\documentclass[twopage,11pt] {article} 
\usepackage{graphicx}
\usepackage{bm}

\usepackage{amssymb,multicol}
\usepackage{amsmath}

\newcommand{\gdhi}{\ooalign{\hfil/\hfil\crcr$\partial$}}
\def\Sp{\mathop{\mathrm{Sp}}\nolimits}
\def\sgn{\mathop{\mathrm{sgn}}\nolimits}

\def\tr{\mathop{\mathrm{tr}}\nolimits}

\begin{document}
\title{Multi-soliton solutions in the chiral quark soliton model}

\author{Nobuyuki Sawado and Noriko Shiiki}

\maketitle

\pagestyle{myheadings} 
\thispagestyle{plain}         
\markboth{Your Name}{N.Sawado and N.Shiiki} 
\setcounter{page}{1}     

In this article a series of solutions with higher baryon numbers
in the chiral quark soliton model are reported.
{\it The chiral quark soliton model} (CQSM) is a simple quark
model that incorporates the basic features of QCD, e.g.
the chiral symmetry and its breakdown accompanied by the appearance of
the Goldstone bosons. It was shown that
the baryon number one ($B=1$) solution provides correct
observable as a nucleon including mass, electromagnetic
value, spin carried by quarks, parton distributions
and octet, decuplet $SU(3)$ baryon spectra.
The $B=2$ axially symmetric soliton solution was obtained numerically
by one of us (N. Sawado). For $B \ge 3$, in a series of our papers 
we obtained $B=3 \sim 9,17$ minimal energy soliton solutions with point-like symmetries 
and $B=3 \sim 5$ axially symmetric saddle-point solutions, using rational map ansatz
constructed for multi-baryon number skyrmions.
An interesting property of the solutions is that the symmetry of the background configuration 
is reflected in the degeneracy of the valence quark spectra.
For instance, the resultant quark bound spectra are doubly degenerate 
for the axisymmetric solitons and are triply degenerate for $B=3$ tetrahedrally 
symmetric solitons.
These results confirm the existence of the quark shell structure.
The shells consist of the four-fold degenerate ground state and higher
levels with various patterns, which are realized by the interplay of two symmetries,
$SU(2)_L\times SU(2)_R$ symmetry for the quarks and the symmetries of the
chiral fields. To obtain correct physical observable, the quantum corrections are
necessary. We shall show the quantum states of the axisymmetric
solitons within the collective quantization. Upon quantization, various
observable spectra of the chiral solitons are obtained.
According to the Finkelstein-Rubinstein constraints,
the quantum numbers of the solitons coincide with the physical observations
only for $B=2$ and $4$ while $B=3$ and 5 do not.
The $SU(3)$ extension of the $B=2$ soliton is also studied to predict
various strange dibaryon states within this model.

\section{\label{sec:level1}Introduction\protect\\ }
 Although QCD is generally accepted as the underlying theory of
the strong interaction, most low- and medium-energy 
nuclear phenomenology may be successfully described in terms of the 
hadronic degrees of freedom. Investigations for deuteron 
photodisintegration and deep inelastic scattering of leptons by 
nuclei suggest the necessity of including  quark degrees 
of freedom~\cite{freedman93,gross92,thomas94,umnikov96,carlson95}.  
For this reason it was suggested that nuclear theory should be reformulated to 
take into account the underlying quark theory. However, QCD is too hard to  
get insight into the low- and medium-energy nuclear phenomenology since  
the coupling constant become large at these energy scales and one can 
not perform perturbation as in the other gauge theories. It is therefore 
necessary to formulate low-energy effective theories for the strong 
interaction.

The chiral quark soliton model is one of such QCD effective theories including 
quark degrees of freedom and baryons as a chiral soliton. 
In $1997$ Diakonov {\it  et al} predicted an exotic 
state with strangeness $+1$ within the topological soliton 
picture of baryons~\cite{diakonov97}, and remarkably such resonance state has been 
experimentally discovered recently~\cite{nakano03}. In a naive point of view, 
this new exotic baryon can be interpreted as a five quark bound state, opening up 
a new paradigm of the nuclei as a multi-quark bound state.    

{\it The chiral quark soliton model} (CQSM) was developed in
1980's as a low-energy effective theory of QCD. 
Since it includes the Dirac sea quark 
contribution and explicit valence quark degrees of 
freedom, the model interpolates between the 
constituent quark model and the Skyrme 
model \cite{diakonov88,reinhardt88,meissner89,report96,wakamatsu91}. 
The CQSM is derived from the instanton liquid model of
QCD vacuum and incorporates the nonperturbative
feature of the low-energy QCD, spontaneous chiral
symmetry breaking. It has been shown that 
the $B=1$ solution provides correct observable as a nucleon 
including mass, electromagnetic value, spin carried
by quarks, parton distributions and octet
$SU(3)$ baryon spectra. 

For $B=2$, the stable axially symmetric
soliton solution was found in Eq.~\cite{sawado98}. 
The solution exhibits doubly degenerate bound spectra  
of the quark orbits in the background of the axially symmetric  
chiral field with winding number two. Upon quantization, 
various dibaryon spectra were obtained, showing that the quantum 
numbers of the ground state coincides with those of a physical 
deuteron~\cite{sawado00,sawado02}. For $B\ge 3$, the Skyrme model 
shows that the soliton solutions have discrete, crystal-like symmetries~\cite{braaten90,sutcliffe97}. 
From the similarity of the chiral field action between the Skyrme model and the CQSM 
one can expect that soliton solutions in the CQSM have same symmetries as skyrmions with 
the same baryon number.
Since it is too complicated to perform a numerical computation if one 
imposes such discrete symmetries directly on the chiral fields,    
Houghton, Manton and Sutcliffe thus proposed remarkable ansatz, rational map ansatz, 
for multi-skyrmions~\cite{manton98}. 
Applying this ansatz to the CQSM we obtained multi-chiral quark soliton solutions   
with point-like symmetries for $B=3\sim 9,17$ as well as saddle-point solutions for 
$B=5,9$~\cite{sawado02t,sawado04}.
The solutions exhibit a large degeneracy and mass gap in the valence quark orbits. 
These results confirm the existence of the quark shell structure.
Esepecially the large degeneracy implies that our solutions with such polyhedral symmetries 
may be the lowest-lying configuration. 

The solitons that we obtain are classical objects and therefore must be 
quantized to assign definite spin and isospin to them.  
Quantization of the solitons can be performed semiclassically 
for their rotational zero modes. 
Quantizing the solutions with discrete symmetries is, however, 
a formidable task in CQSM. 
Thus, before embarking those discrete symmetries, 
it will be instructive to study axially symmetric 
solutions which are much simpler~\cite{komori04}. Besides, considering 
the fact that for some higher baryon numbers, the ground 
states of the skyrmions do not agree with the experimental 
observation~\cite{irwin} , the possibility that axially symmetric 
solutions may provide correct ground states can not be 
excluded. In fact it was found in Ref.~\cite{grigoriev} that
the axially symmmetric BPS monopoles up to 
chrage five have lower energies than those of discrete symmetries. 
We therefore investigate classical and quantum multi-soliton solutions in the CQSM 
with axial symmetry up to $B=5$. 

Since the first prediction of the H-particle in a MIT bag model 
calculation~\cite{jaffe77}, there have been many efforts to 
study the spectrum of dibaryonic systems including strangeness. 
We shall apply our formulation to study these six quark states constituting 
a dibaryon and make a prediction for its mass spectra.  

In Sec.~\ref{sec:level2} the formulation of the 
chiral quark soliton model is introduced. 
In Sec.~\ref{sec:level3} we obtain axially symmetric soliton solutions for 
$B=2\sim 5$ and solutions with the polyhedral symmetries for $B=3\sim 9$ 
including saddle-point configurations. 
The relation between the symmetries of the chiral fields and the 
degeneracy of the valence quark spectra  
is discussed in Sec.~\ref{sec:level4} from a group theoretical point of view.  
In Sec.~\ref{sec:level5} we perform zero mode quantization for the obtained classical 
solitons. Imposing the Finkelstein-Rubinstein 
constraints on the states, the ground states of the axially symmetric 
solitons are constructed and examined if they agree with the 
experimental observation. 
Conclusions and discussions are in Sec.\ref{sec:level6}. 
 
\section{\label{sec:level2}The Chiral Quark Soliton Model : General Formalism}
The CQSM is derived from the instanton liquid model of 
the QCD vacuum and incorporates the nonperturbative feature
of the low-energy QCD, spontaneous chiral symmetry breaking. 
The vacuum functional is defined by Ref.~\cite{diakonov88}
\begin{eqnarray}
	{\mathcal Z} = \int {\mathcal D}\pi{\mathcal D}\psi{\mathcal D}\psi^{\dagger}
	\exp \left[i \int d^{4}x \, \bar{\psi} \left(i\!\!\not\!\partial
	- MU^{\gamma_{5}}\right) \psi \right]	 \label{vacuum_functional}
\end{eqnarray} 
where the SU(2) matrix
\begin{eqnarray}
	U^{\gamma_{5}}= \frac{1+\gamma_{5}}{2} U + \frac{1-\gamma_{5}}{2} U^{\dagger} 
	\,\,\,{\rm with} \,\,\,\,
	U=\exp \left( i \bm{\tau} \!\cdot\! \bm{\pi}/f_{\pi} \right) \nonumber
\end{eqnarray}
describes chiral fields, $\psi$ is quark fields and $M$ is the constituent 
quark mass. $f_{\pi} $ is the pion decay constant and experimentally 
$f_{\pi} \sim 93 {\rm MeV}$. 

The $B=1$ soliton solution has been studied in detail at classical and 
quantum level in Refs.~\cite{diakonov88,reinhardt88,meissner89,report96,wakamatsu91}.  
To obtain solutions with $B>1$, we shall employ the chiral fields with 
winding number $B$ in the Skyrme model as the background of quarks, 
which can be justified as follows. 

In Eq.~(\ref{vacuum_functional}), performing the functional integral over 
$\psi$ and $\psi^\dagger$ fields, one obtains the effective action
\begin{equation}
	S_{\rm eff}(U)=-iN_{c} {\rm Sp}~{\rm ln}  iD = -iN_c \log \det iD, 
	\label{effective_action2}
\end{equation}
where $iD=i \gdhi - M U^{\gamma_5}$ is the Dirac operator.
The classical solutions can be obtained by the extremum condition 
of (\ref{effective_action2}) with respect to $U$. 
For this purpose, let us consider the derivative expansion of 
the action \cite{dhar85,ebert86,wakamatsu91}. 
Up to quartic terms, we have,  
\begin{eqnarray}
	S_{\rm eff}(U)=\int d^4x \biggl[-{\mathcal C} {\rm tr}(L_\mu L^\mu)
	+\frac{N_c}{32\pi^2}
	{\rm tr}\Bigl\{\frac{1}{12}[L_\mu,L_\nu]^2-\frac{1}{3}(\partial_\mu L^\mu)^2
	+\frac{1}{6}(L_\mu L^\mu)^2 \Bigr\} \biggr], 
	\label{expand_action}
\end{eqnarray}
where $L_\mu=U^\dagger \partial_\mu U$.
The baryon number $B$ can be calculated in terms of the topological charge, 
\begin{eqnarray}
	B=-\frac{1}{24\pi^2}\epsilon_{ijk}\int d^3x{\rm tr}(L_iL_jL_k)\,.
	\label{wess_zumino}
\end{eqnarray} 
Suitably adjusting the coefficients ${\mathcal C}$, 
one can identify the first two terms of Eq.~(\ref{expand_action}) with the Skyrme 
model action. However, the 4th order terms tend to destabilize solutions and  
no stable classical solution can be obtained from the above action~\cite{dhar85,aitchison85}. 
Nevertheless, because of their similarity, it will be justified to adopt the 
configurations of the solutions in the Skyrme model to chiral fields in the CQSM. 

In the CQSM, the number of valence quark is associated with 
the baryon number such that the baryon number $B$ soliton consist of 
$N_c\times B$ valence quarks. 
If the quarks are strongly bound inside the soliton, 
their binding energy become large and the valence quarks 
can not be observed as positive energy particles
~\cite{kahana84,balachandran98}.
Thus, one gets the picture of the topological soliton model 
in the sense that the baryon number coincide with the winding 
number of the background chiral field when the valence quarks 
occupy all the levels diving into negative energy region. 

Let us rewrite the effective action in Eq.~(\ref{effective_action2}) as 
\begin{eqnarray}
	S_{\rm eff}= -iN_c \log \det(i \gdhi - M U^{\gamma_5}) 
	=-iN_c\log \det\bigl(i\partial_t-H(U^{\gamma_5})\bigr) 
	\label{effective_det}
\end{eqnarray}
where 
\begin{eqnarray}
	H(U^{\gamma_5})=-i\alpha\cdot\nabla + \beta MU^{\gamma_{5}}\,.
	\label{hamiltonian}
\end{eqnarray}
The classical energy of the soliton can be estimated from the quark determinant
in Eq.~(\ref{effective_det}) \cite{rajaraman,reinhardt89}. 
We introduce the eigenstates of operators, 
$i\partial_t-H(U^{\gamma_5})$ and $H(U^{\gamma_5})$, such that 
\begin{eqnarray}
	&&H(U^{\gamma_5})\phi_{\mu}(\bm{x})=E_{\mu}\phi_{\mu}(\bm{x})\,, \label{eigen_h}\\
	&&\bigl(i\partial_t-H(U^{\gamma_5})\bigr)\Psi_{\mu,n}=\lambda_{\mu,n}\Psi_{\mu,n}\,,
	\label{eigenequation}
\end{eqnarray}
where $\Psi_{\mu,n}=e^{-i\omega_{n}t}\phi_{\mu}$ and 
$\lambda_{\mu,n}=-E_\mu+\omega_n$.
Imposing on $\Psi_{\mu,n}$ the anti-periodicity condition,  
$\Psi_{\mu,n}(\bm{x},T)=-\Psi_{\mu,n}(\bm{x},0)$, 
reads  
\begin{eqnarray}
\omega_nT=(2n+1)\pi. 
\end{eqnarray}
The determinant in Eq.~(\ref{effective_det}) then becomes 
\begin{eqnarray}
	\det(i\partial_t-H)&=&\prod_{\mu,n}\lambda_{\mu,n} 	
	=\prod_{\mu,n}\Bigl(-E_\mu+\frac{(2n+1)\pi}{T}\Bigr) \nonumber \\
	&=&C\prod_{\mu,n\ge0}\Bigl(1-\frac{|E_\mu|^2 T^2}{(2n+1)^2\pi^2} \Bigr) 
	=C\prod_{\mu}\cos\Bigl(\frac{1}{2}|E_\mu|T\Bigr) \nonumber \\ 
	&=&\frac{C}{2}\exp\Bigl(i\frac{1}{2}\sum_\mu|E_\mu|T\Bigr)
	\prod_\mu\Bigl(1+\exp(-i|E_\mu|T)\Bigr) \label{det}
\end{eqnarray}
where 
\begin{equation}
	C=\prod_{n\ge0}\Bigl(-\frac{(2n+1)^2\pi^2}{T^2}\Bigr) \nonumber
\end{equation}
and the product formula for the cosine function 
$\cos(z)=\prod^\infty_{n\ge1}(1-4z^2/(2n-1)^2\pi^2)$ has been used.
Inserting (\ref{det}) into Eq.~(\ref{effective_det}), one obtains 
\begin{eqnarray}
	S_{\rm eff}=-N_cT\sum_\mu n_\mu|E_\mu|+N_cT\frac{1}{2}\sum_\mu |E_\mu|\,,
\end{eqnarray}
where $n_\mu$ is the valence quark occupation number which takes  
values only 0 or 1. 
Correspondingly, the classical energy is given by
\begin{eqnarray}
	E_{\rm static}=E_{\rm val}+E_{\rm vac}
\end{eqnarray}
where
\begin{eqnarray}
	E_{\rm val}=N_c\sum_\mu n_\mu |E_\mu|\,,\,\, 
	E_{\rm vac}=-\frac{1}{2}N_c\sum_\mu |E_\mu|\,,\nonumber
\end{eqnarray}
representing the valence quark and 
sea quark contribution to the total energy respectively. 

The effective action $S_{\rm eff}(U)$ is ultraviolet divergent 
and hence must be regularized. 
Using the proper-time regularization scheme 
\cite{schwinger51}, we can write 
\begin{eqnarray}
	S^{{\rm reg}}_{{\rm eff}}[U]
	&=&\frac{i}{2}N_{c}
	\int^{\infty}_{1/\Lambda^2}\frac{d\tau}{\tau}{\rm Sp}\left(
	{\rm e}^{-D^{\dagger}D\tau}-{\rm e}^{-D_{0}^{\dagger}D_{0}\tau}\right) 
	\nonumber \\
	&=&\frac{i}{2}N_{c}T\int^{\infty}_{-\infty}\frac{d\omega}{2\pi}
	\int^{\infty}_{1/\Lambda^2}\frac{d\tau}{\tau}
	{\rm Sp}\Bigl[{\rm e}^{-\tau (H^2+\omega^2)} 
	-{\rm e}^{-\tau (H_{0}^2+\omega^2)}\Bigr]
	\label{regularized_action}
\end{eqnarray} 
where $D_{0}$ and $H_{0}$ are operators with $U=1$.
The total energy is then given by
\begin{equation}
	E_{\rm static}[U]=E_{\rm val}[U]+E_{\rm vac}[U]-E_{\rm vac}[U=1]
	\label{total_energy}
\end{equation}
where
\begin{eqnarray}
	E_{\rm val}=N_c\sum_{i}E^{(i)}_{\rm val}\,,\,\,
	E_{\rm vac}=N_c\sum_{\mu}\left\{ {\mathcal N}(E_\mu)|E_{\mu}|+\frac{\Lambda}
	{\sqrt{4\pi}}\exp \left[ - \left( \frac{E_{\mu}}{\Lambda} \right)^2 
	\right] \right\}\nonumber 
\end{eqnarray}
with
\begin{equation}
	{\mathcal N}(E_{\mu})= -\frac{1}{\sqrt{4\pi}}\Gamma \left(\frac{1}{2}, 
	\left(\frac{E_{\mu}}{\Lambda} \right)^2 \right) \,\nonumber
\end{equation}
and $E^{(i)}_{\rm val}$ is the valence energy of the $i$ th 
valence quark. $\Lambda$ is a cutoff parameter evaluated by
the condition that the derivative expansion of Eq.~(\ref{regularized_action}) 
reproduces the pion kinetic term with the correct coefficient, {\it i.e.},
\begin{eqnarray}
	f_{\pi}^2=\frac{N_{c}M^2}{4\pi^2}\int^{\infty}_{1/\Lambda^2} 
	\frac{d\tau}{\tau}{\rm e}^{-\tau M^2}
	\, . \label{cutoff_parameter}
\end{eqnarray}
In this model, the constituent quark mass $M$ is the only free parameter 
and we take the value $M=400$ MeV, in which the observable of the nucleon 
and the delta are well reproduced~\cite{report96}. From Eq.~(\ref{cutoff_parameter}) 
and by using the values of $M,f_\pi$, we obtain $\Lambda\sim 637$ MeV. 

For $B=1$, one imposes a spherically symmetric ansatz ({\it hedgehog ansatz}) 
\begin{eqnarray}
U(\bm{r})=\exp(i F(r) \hat{\bm{r}}\cdot \bm{\tau})=\cos F(r)+i\hat{\bm{r}}\cdot \bm{\tau}\sin F(r)\,,
\label{chiral_fields_hedgehog}
\end{eqnarray}
with the boundary condition for the profile function $F(r)$
\begin{eqnarray}
F(0)=-\pi,~~F(\infty)=0\,.
\label{boundary_condition}
\end{eqnarray}
Substituting the ansatz (\ref{chiral_fields_hedgehog}) into Eq.~(\ref{wess_zumino}) 
under the boundary condition (\ref{boundary_condition}), one gets 
\begin{eqnarray}
  B=\frac{1}{2\pi^2}\int^\infty_0 \frac{\sin^2F(r)}{r^2}\frac{dF(r)}{dr}4\pi r^2dr
  =\frac{1}{\pi}\left[F+\frac{\sin 2F}{2}\right]_{-\pi}^{0}=1\,.
\end{eqnarray}
The one-quark hamiltonian (\ref{hamiltonian}) becomes  
\begin{eqnarray}
H(U^{\gamma_5})=-i\alpha\cdot\nabla + \beta M(\cos F(r)+i\gamma_5\hat{\bm{r}}\cdot 
\bm{\tau}\sin F(r))\,.
	\label{hamiltonian_hedgehog}
\end{eqnarray}
This hamiltonian does not commute with the total angular momentum $\bm{J}$ nor the 
isospin $\bm{\tau}/2$ but commute with the grand spin operator $\bm{K}=\bm{J}+\bm{\tau}/2$. 
$H$ also commutes with the parity operator ${\cal P}=\gamma_0$.  
Hence the one-quark eigenstates are labeled by the $K=0,1,2,\cdots$ and the 
parity ${\cal P}=\pm$. The three valence quarks occupy their 
lowest states $K^{\cal P}=0^{+}$ and responsible for the baryon number of the soliton $(=1)$.  
In this context, the baryon number is not a topological origin unlike the Skyrme @model. 
 
Field equations for the chiral fields can be obtained by demanding  
that the total energy in Eq.~(\ref{total_energy}) be stationary 
with respect to variation of the profile function $F(r)$,
\begin{eqnarray*}
	\frac{\delta}{\delta F(r)}E_{\rm static}=0 \,\, ,
\end{eqnarray*}
which produces  
\begin{eqnarray}
	S(r)\sin F(r)=P(r)\cos F(r),  
	\label{field_equation}
\end{eqnarray}
where 
\begin{eqnarray}
&&S(r)=N_{c}\sum_\mu\bigl(n_\mu\theta(E_\mu)+{\rm sign}(E_\mu)
{\mathcal N}(E_\mu)\bigr)
\langle \mu |\gamma^{0}\delta(|x|-r)|\mu\rangle\,, 
\\	
&&P(r)=N_{c}\sum_\mu\bigl(n_\mu\theta(E_\mu)+{\rm sign}(E_\mu)
{\mathcal N}(E_\mu)\bigr)
\langle \mu |i \gamma^{0}\gamma^{5}\hat{\bm{r}}
\cdot\bm{\tau}\delta(|x|-r)|\mu\rangle \, .
\end{eqnarray}
The procedure to obtain self-consistent solutions of Eq.~(\ref{field_equation}) 
is that $1)$ solve the eigenequation in the hamiltonian (\ref{hamiltonian_hedgehog}) 
under an assumed initial profile function $F_{0}(r)$, $2)$ use the resultant eigenfunctions and 
eigenvalues to calculate $S(r)$ and $P(r)$, $3)$ solve 
(\ref{field_equation}) to obtain a new profile function, $4)$ repeat $1)-3)$ 
until the self-consistency is attained.

The calculated energy of the $B=1$ soliton is $E_{\rm static}=1192$ MeV with the 
constituent quark mass $M=400$ MeV. 

\section{\label{sec:level3}The Classical Configurations}
\subsection{The Axially Symmetric Configuration}
It is known that the minimal energy solution for $B>1$ is not spherically symmetric.  
In the Skyrme model the configuration with $B=2$ is axially  
symmetric \cite{manton87,kopeliovich87,verbaashot87,weigel86} 
and can be written by 
\begin{equation}
	U(\bm{x})= \cos F(\rho,z)+i\bm{\tau} \cdot \hat{\bm{n}} \sin F(\rho,z),
	\label{ansatz}
\end{equation}
where
\begin{equation}
	\hat{\bm{n}}=(\sin \Theta(\rho,z) \cos m_{{\rm w}} \varphi,\sin \Theta(\rho,z) 
	\sin m_{{\rm w}} \varphi,\cos \Theta(\rho,z))
	\label{ansatz2}
\end{equation}
and $m_{{\rm w}}$ is the winding number of the pion fields. We shall use this 
configuration in the background to obtain axially symmetric chiral quark 
solitons.

The extremum conditions for the total energy 
\begin{equation}
\frac{\delta}{\delta F(\rho,z) } E_{\rm static}[U]=0\,,~~~\frac{\delta}{\delta 
\Theta(\rho,z) } E_{\rm static}[U]=0
\end{equation}
yield the following equations of motion for the profile functions,
\begin{eqnarray}
	&&R^{T}(\rho,z)\cos \Theta (\rho,z)=R^{L}(\rho,z)\sin \Theta (\rho,z)\,,
	\label{eq_proff}  \\
	&&S(\rho,z)\sin F (\rho,z)=P(\rho,z)\cos F (\rho,z)~\label{eq_proft}
	\end{eqnarray}
where
\begin{equation}
	P(\rho,z)=R^{T}(\rho,z)\sin \Theta (\rho,z)+R^{L}(\rho,z)\cos \Theta (\rho,z)\,.
	\label{eq_p}
\end{equation}
In terms of eigenfunction $\phi$ in Eq.~(\ref{eigen_h}), 
$R^{T}$,$R^{L}$ and $S$ are given by 
\begin{eqnarray}
	&&R^{T}(\rho,z)=R^{T}_{\rm val}(\rho,z)+R^{T}_{\rm vac}(\rho,z)\,, \label{eq_rt}\\
	&&R^L(\rho,z)=R^{L}_{\rm val}(\rho,z)+R^L_{\rm vac}(\rho,z)\,,  \label{eq_rl} \\
	&&S(\rho,z)=S_{\rm val}(\rho,z)+S_{\rm vac}(\rho,z) \label{eq_s}
\end{eqnarray}
where
\begin{eqnarray}
	&&R^{T}_{\rm val}(\rho,z)=\sum_{i}\int d \varphi \bar{\phi}_i(\rho,\varphi,z)
	 i \gamma_5
	(\tau_1 \cos m_{{\rm w}}\varphi + \tau_2 \sin m_{{\rm w}}\varphi ) 
	\phi_i(\rho,\varphi,z)\,, \nonumber\\
	&&R^{T}_{\rm vac}(\rho,z)= \sum_{\mu} {\mathcal N}(E_{\mu}) \sgn(E_{\mu}) \int d 
	\varphi \bar{\phi}_{\mu}(\rho,\varphi,z)
	 i \gamma_5\nonumber\\
	&&\hspace{2cm}\times
	(\tau_1 \cos m_{{\rm w}}\varphi + \tau_2 \sin m_{{\rm w}}\varphi ) 
	\phi_{\mu}(\rho,\varphi,z)\,, \nonumber\\
	&&R^L_{\rm val}(\rho,z)=\sum_{i}\int d \varphi \bar{\phi}_i(\rho,\varphi,z)i 
	\gamma_5\tau_3\phi_i(\rho,\varphi,z)\,, \nonumber\\
	&&R^L_{\rm vac}(\rho,z)= \sum_{\mu} {\mathcal N}(E_{\mu}) \sgn(E_{\mu}) \int d \varphi 
	\bar{\phi}_{\mu}(\rho,\varphi,z)
	 i \gamma_5
	\tau_3 \phi_{\mu}(\rho,\varphi,z)\,, \nonumber\\
	&&S_{\rm val}(\rho,z)=\sum_{i}\int d \varphi \bar{\phi}_i(\rho,\varphi,z)
	\phi_i(\rho,\varphi,z)\,,  \nonumber\\
	&&S_{\rm vac}(\rho,z)= \sum_{\mu} {\mathcal N}(E_{\mu}) \sgn(E_{\mu}) \int d \varphi 
	\bar{\phi}_{\mu}(\rho,\varphi,z)\phi_{\mu}(\rho,\varphi,z)\,,\nonumber 
\end{eqnarray}
and subscripts, vac and val, represent the vacuum and valence quark 
contributions respectively.
The boundary conditions for the profile functions were constructed 
by Braaten and Carson~\cite{braaten88};
\begin{eqnarray}
	&&F(\rho ,z)\rightarrow 0 ~~{\rm as}~~ \rho^2+z^2\rightarrow \infty , \nonumber \\
	&&F(0,0)=-\pi, ~~~\Theta(0,z)=  \left\{ 
	\begin{array}{c} 0, ~~~ z>0  \\
	\pi, ~~~z<0  
	\end{array}\right.\;.
	\label{boundary}
\end{eqnarray}

In Fig.~\ref{fig:Sf}, we show the spectrum of the 
quark orbits in the background of chiral fields with winding number 
$m_{{\rm w}}=2,4$, as a function of the size parameter $X$.
The axially symmetric profile functions are parameterized by $X$ as

\begin{eqnarray}
	F(\rho,z)&=&-\pi+\pi\sqrt{\rho^2+z^2}/X~~{\rm for}~~\sqrt{\rho^2+z^2}\le X 
	\nonumber \\
	&=&0\hspace{3.1cm}{\rm otherwise}, \\
	\Theta(\rho,z)&=&\tan^{-1}(\rho/z).
\end{eqnarray}  
To examine the spectrum in detail, let us consider the hamiltonian 
defined in Eq.~(\ref{hamiltonian}).
For the axially symmetric chiral field in Eq.~(\ref{ansatz}), this 
hamiltonian commutes with the third component of the grand spin 
operator $K_{3}$ and the time-reversal operator $T$. 
These are specifically, 
\begin{eqnarray}
 	 &&K_3=L_3+\frac{1}{2} \sigma_3 +\frac{1}{2} m_{{\rm w}}\tau_3\,, \\
	 &&T=i \gamma_1 \gamma_3 \cdot i \tau_1 \tau_3 C
\end{eqnarray}
where $L_3$, $\sigma_3$, and $\tau_3$ are respectively 
the third component of orbital angular momentum, spin, and isospin operator,  
and $C$ is a charge conjugation operator. 
The parity operator is defined by $P=\gamma_0$ for odd $B$, 
and $P=\gamma_0\tau_3$ for even $B$.
Thus, the eigenvalues of the hamiltonian can be specified 
by the magnitude of $K_3$ and the parity $\pi=\pm$. 
We have $K_3 = 0, \pm1, \pm2, \pm3,\cdots $ for odd $B$, 
and $K_{3}=\pm\frac{1}{2}, \pm\frac{3}{2}, 
\pm\frac{5}{2}, \pm\frac{7}{2}, \cdots $ for even $B$.  
Since the hamiltonian 
is invariant under time reverse, the states of 
$+K_3$ and $-K_3$ are degenerate in energy. 

From Fig.~\ref{fig:Sf} it can be seen that for $m_{{\rm w}}=2$, the bound states diving into negative 
region are doubly degenerate with $K_3=\pm\frac{1}{2}$. 
Thus putting $N_c=3$ valence quarks on each of the bound levels, we have
the $B=2$ soliton solution. 
Similary, for $m_{{\rm w}}=3$, we obtained the spectrum of $K_3 = \pm 1^{-}$(double degeneracy) and
 $K_3 = 0^{+} $ states diving into negative-energy region which corresponds to the $B=3$ soliton solution. 
For $m_{{\rm w}}=4$, the spectrum of $K_3 = \pm {\frac{1}{2}}^{+}$ 
and $K_3 = \pm {\frac{3}{2}}^{-}$(both doubly degenerate) states 
dive into negative region, having the $B=4$ soliton solution.
For $m_{{\rm w}}=5$, the spectrum of $K_3 = \pm 2^{+}$(double), 
$K_3 = \pm 1^{-}$(double) and $K_3 = 0^{+} $ states dive into 
negative-energy region, having the $B=5$ soliton solution. 
These results confirm that the baryon number of the soliton is identified 
with the number of diving levels occupied by $N_c$ valence quarks.
It is interesting that the degeneracy which occurs due to symmetry of 
the chiral field reduces the number of states, making large shell gaps. 
This observation indicates that degeneracy in the valence quark spectra 
contributes to make classical energies of the soliton solutions lower. 
In fact, our $B=2$ solution which is considered to be the minimum energy 
soliton from the study of the $B=2$ skyrmion provides the maximum degeneracy 
in spectra. It will be interesting to study minimum solutions from this point 
of view. 

\begin{table}
\begin{center}
\caption{\label{tab:claspectra}The classical observable, Mass (in MeV), 
the mean radius of toroid and the root mean square radius (in fm) for $B=1\sim 5$.}
\begin{tabular}{cccccccccc}\hline 
$B$ & \multicolumn{5}{c}{$E_{\rm val}$} & $E_{\rm vac}$ & 
$E_{\rm static}$ &$\langle \rho \rangle$  &$\sqrt{\langle r^2 \rangle}$ \\ \hline 
1 & 173 &     &     &     &     & 674  & 1192 &       & 0.785 \\
2 & 173 & 173 &     &     &     & 1166 & 2204 & 0.672 & 0.821 \\
3 & 173 & 173 & 298 &     &     & 1561 & 3493 & 0.659 & 0.854 \\
4 & 106 & 106 & 232 & 232 &     & 2727 & 4753 & 0.971 & 1.140 \\
5 & 145 & 145 & 319 & 319 & 409 & 2537 & 6543 & 1.048 & 1.225 \\ \hline
\end{tabular}
\end{center}
\end{table}

The baryon number density is defined by the zeroth component of 
the baryon current \cite{reinhardt88}, 
\begin{equation}
	b(\bm{x}) =\frac{1}{N_c}\langle \bar{\psi}\gamma_0 \psi \rangle
	=b_{\rm val}(\bm{x})+b_{\rm vac}(\bm{x})
	\label{bdensity_axial}
\end{equation}
where
\begin{eqnarray}
	b_{\rm val}(\bm{x})&=&\frac{1}{N_{c}}\sum_{i}\int d\varphi\, 
	\phi_{i}^{\dagger}(\rho,\varphi,z)\phi_{i}(\rho,\varphi,z) \nonumber\\
	b_{\rm vac}(\bm{x})&=&\frac{1}{N_{c}}\biggl[
	\sum_{\mu}{\mathcal N}(E_{\mu})\sgn (E_{\mu})
	\int d\varphi \, \phi_{\mu}^{\dagger}
	(\rho,\varphi,z)\phi_{\mu}(\rho,\varphi,z) 
	\nonumber \\
	&-&\sum_{\mu}{\mathcal N}(E_{\mu}^{(0)})
	\sgn (E_{\mu}^{(0)})  
	\int d\varphi \phi_{\mu}^{(0)\dagger}(\rho,\varphi,z)
	\phi_{\mu}^{(0)}(\rho,\varphi,z) \biggr]\,.\nonumber
\end{eqnarray}
A contour plot of the baryon number density for each baryon number
is shown in Fig.~\ref{fig:Bd}. It can be seen that they have toroidal 
in shape. 

The mean radius ${\langle \rho \rangle}$ is given by 
\begin{equation}
	\langle \rho \rangle ={\langle \rho \rangle}_{\rm val}
	+{\langle \rho \rangle}_{\rm vac}
\end{equation}
where
\begin{eqnarray*}
	&&{\langle \rho \rangle}_{\rm val} = \frac{1}{m_{{\rm w}}}
	\sum_i \int \rho d \rho d z d 
	\varphi \rho  \phi_{i}^{\dagger} (\rho,\varphi,z)
	\phi_{i}(\rho,\varphi,z)\,,\\
	&&{\langle \rho \rangle}_{\rm vac} = \frac{1}{m_{{\rm w}}} 
	\sum_\mu {\mathcal N}(E_\mu) 
	\sgn(E_{\mu})  
	\int \rho d \rho d z d \varphi 
	\rho \phi_{\mu}^{\dagger} 
	(\rho,\varphi,z)\phi_{\mu}(\rho,\varphi,z)\,.
\end{eqnarray*}
The root mean square radius is given by 
\begin{equation}
	\sqrt{\langle r^2 \rangle} =\sqrt{\langle r^2 \rangle_{\rm val}+
	\langle r^2 \rangle_{\rm vac}}
\end{equation}
where
\begin{eqnarray}
	&& \langle r^2 \rangle_{\rm val} = \frac{1}{m_{{\rm w}}}\sum_i \int \rho d \rho d z d 
	\varphi ({\rho}^2 + z^2) \phi_{i}^{\dagger} (\rho,\varphi,z)\phi_{i}
	(\rho,\varphi,z)\,,  \nonumber \\
	&&\langle r^2 \rangle_{\rm vac} = \frac{1}{m_{{\rm w}}} \sum_\mu {\mathcal N}(E_\mu) 
	\sgn(E_{\mu})
	\int \rho d \rho d z d \varphi ({\rho}^2 + z^2) 
	\phi_{\mu}^{\dagger} (\rho,\varphi,z)\phi_{\mu}(\rho,\varphi,z)\,.\nonumber 
\end{eqnarray}
These values for each baryon number are shown in Table~\ref{tab:claspectra}.  

\subsection{Multi-Winding Number Configurations with Polyhedral Symmetries}
In the Skyrme model it is known that minimal energy configurations with $B\ge 3$ have 
discrete crystal-like symmetries~\cite{braaten90} rather than axisymmetry. 
We expect that configurations of the CQSM inherits the same discrete 
symmetry as skyrmions. 
However, it is too complicated to perform a numerical computation if one 
imposes such discrete symmetries directly on the chiral fields. Thus  
Houghton, Manton and Sutcliffe proposed remarkable ansatz for the chiral fields, 
rational map ansatz~\cite{manton98}. 
According to this ansatz, the chiral fields are expressed in a rational map as 
\begin{eqnarray}
U(r,z)=\exp(i F(r) \hat{\bm{n}}_R\cdot \bm{\tau})\,,
\label{chiral_fields_RMA}
\end{eqnarray}
where
\begin{eqnarray}
\hat{\bm{n}}_R=\frac{1}{1+|R(z)|^2}(2{\rm Re}[R(z)],2{\rm Im}[R(z)],1-|R(z)|^2)
\nonumber
\end{eqnarray}
and $R(z)$ is the rational map. The complex coordinate $z$ is given by  
$z=\tan (\theta/2)e^{i\varphi}$ via stereographic projection. 

Rational maps are maps from $CP(1)$ to $CP(1)$ (equivalently, from $S^2$ 
to $S^2$) classified by winding number. In Ref.~\cite{manton98} Houghton {\it et al.} 
showed that $B=N$ skyrmions can be well-approximated by rational maps with 
winding number $N$. The rational map with winding number $N$ possesses 
$(2N+1)$ complex parameters whose values can be determined so as to realize 
minimal energies within assumed symmetries of the skyrmion. We shall use this ansatz for 
the background chiral fields in the CQSM. 
Their explicit forms of the map are~\cite{manton98,battye01}
\begin{eqnarray}
&&R_3=\frac{\sqrt{3}iz^{2}-1}{z(z^{2}-\sqrt{3}i)} \,,\nonumber\\
&&R_4=\frac{z^4+2\sqrt{3}iz^2+1}{z^4-2\sqrt{3}iz^2+1} \,,\nonumber \\
&&R_5=\frac{z(z^4+3.94z^2+3.07)}{3.07z^4-3.94z^2+1} \,, \nonumber \\
&&R_6=\frac{z^4+0.16 i}{z^2(0.16z^4 i+1)} \,,\nonumber \\
&&R_7=\frac{7/\sqrt{5}z^6-7z^4-7/\sqrt{5}z^2-1}{z(z^6+7/\sqrt{5}z^4+7z^2-7/\sqrt{5})} \,,\nonumber \\
&&R_8=\frac{z^6-0.14}{z^2(0.14z^6+1)}\,, \nonumber \\
&&R_9=\frac{z(-3.38-11.19iz^4+z^8)}{1-11.19iz^4-3.38z^8}\,,\nonumber \\
&&R_5^{*}=\frac{z(z^4-5)}{-5z^4+1} \,,\nonumber \\
&&R_9^{*}=\frac{5i\sqrt{3}z^6-9z^4+3i\sqrt{3}z^2+1-1.98z^2(z^6-i\sqrt{3}z^4-z^2+i\sqrt{3})}
{z^3(-z^6-3i\sqrt{3}z^4+9z^2-5i\sqrt{3}-1.98z(-i\sqrt{3}z^6+z^4+i\sqrt{3}z^2-1))}\,, \nonumber \\
&&R_{17}=\frac{17z^{15}-187z^{10}+119z^5-1}{z^2(z^{15}+119z^{10}+187z^5+17)}\,.
\end{eqnarray}
Field equations for the chiral fields can be obtained by demanding 
that the total energy in Eq.~(\ref{total_energy}) be stationary 
with respect to variation of the profile function $F(r)$,
\begin{eqnarray*}
	\frac{\delta}{\delta F(r)}E_{\rm static}=0 \,\, ,
\end{eqnarray*}
which produces  
\begin{eqnarray}
	S(r)\sin F(r)=P(r)\cos F(r),  
	\label{field_equation_RMA}
\end{eqnarray}
where 
\begin{eqnarray}
&&S(r)=N_{c}\sum_\mu\bigl(n_\mu\theta(E_\mu)+{\rm sign}(E_\mu)
{\mathcal N}(E_\mu)\bigr)
\langle \mu |\gamma^{0}\delta(|x|-r)|\mu\rangle\,, 
\\	
&&P(r)=N_{c}\sum_\mu\bigl(n_\mu\theta(E_\mu)+{\rm sign}(E_\mu)
{\mathcal N}(E_\mu)\bigr)
\langle \mu |i \gamma^{0}\gamma^{5}\hat{\bm{n}}_{R}
\cdot\bm{\tau}\delta(|x|-r)|\mu\rangle \, .
\end{eqnarray}
\begin{table}
\begin{center}
\caption{\label{tab:energy} Mass spectra for $B=1-9,17$ also for some excited states
$B=5^{*},9^{*}$(in MeV).
The data for $B=2$ are taken from Ref.~\cite{sawado00}.
The ratio of the mass $E_{\rm static}$ to $B\times E^{(B=1)}_{\rm static}$ are compared to 
that of the Skyrme model~\cite{manton98}.}
\begin{tabular}{lccccccccccccc} \hline
$B$ &\multicolumn{9}{c}{$E^{(i)}_{\rm val}$}&$E_{\rm field}$ 
&$E_{\rm static}$ &\multicolumn{2}{c}{$E_{\rm static}/B E^{(B=1)}_{\rm static}$}\\
    &\multicolumn{9}{c}{}  & && Ours & Skyrme\\ \hline
1    & 173 &     &     &     &     &&&  &  & 674  & 1192 & 1.00 & 1.00\\
2    & 173 & 173 &     &     &     &&&  &  & 1166 & 2204 & 0.92 & 0.95\\     
3    & 210 & 210 & 210 &     &     &&&  &  & 1633 & 3522 & 0.98 & 0.96\\ 
4    & 144 & 146 & 146 & 146 &     &&&  &  & 2628 & 4378 & 0.92 & 0.92\\
5    & 123 & 131 & 131 & 139 & 210 &&&  &  & 3265 & 5467 & 0.92 & 0.93\\ 
6    & 120 & 124 & 150 & 150 & 206 & 206 &&&     & 3740 & 6603 & 0.92 &0.92   \\
7    & 115 & 120 & 120 & 120 & 166 & 166 & 166 &&& 4554 & 7478 & 0.90 & 0.90 \\
8    & 97  & 97  & 115 & 120 & 139 & 139 & 203 & 203 && 5229 & 8565 & 0.90 & 0.91 \\
9    & 69  & 101 & 104 & 104 & 107 & 166 & 166 & 179 & 179 & 6046 & 9573 & 0.89& 0.90{\tiny 6} \\
17   & 83  & 95  & 95  & 95  & 153 & 156 & 157 & 173 & 175 &      &      &     &      \\
     & 177 & 178 & 179 & 192 & 194 & 194 & 196 & 196 &     & 10586& 18650& 0.93& 0.88\\
$5^{*}$& 157 & 157 & 157 & 232 & 232 &&&  &  & 2874 & 5680 & 0.95& 1.00\\ 
$9^{*}$& 99  & 105 & 105 & 121 & 142 & 142 & 210 & 210 & 210 & 5700 & 9742 & 0.91& 0.91 \\
\hline
\end{tabular}
\end{center}
\end{table}

The baryon density $b(\bm{x})$ is defined by 
\begin{eqnarray}
	&&b(\bm{x}) = \frac{1}{N_{c}}\langle\bar{\psi}\gamma_{0}
	\psi\rangle=b_{\rm val}(\bm{x})+b_{\rm field}(\bm{x}),
\label{baryon_density}
\end{eqnarray}
where
\begin{eqnarray}
	b_{\rm val}(\bm{x})&=& \sum_i b_{\rm val}^{(i)}(\bm{x})
	=\frac{1}{N_{c}}
	\sum_{i}  
	\phi_i(\bm{x})^\dagger \phi_i(\bm{x})\,,
	\nonumber \\
	b_{\rm field}(\bm{x})&=&\frac{1}{N_c}\Bigl
	[\sum_{\mu}{\rm sign}(E_\mu){\mathcal N}(E_\mu)	
	\phi_\mu(\bm{x})^\dagger \phi_\mu(\bm{x})\nonumber \\
	&-&\sum_{\mu}{\rm sign}(E_\mu^{(0)}){\mathcal N}(E_\mu^{(0)})	
	\phi_\mu^{(0)}(\bm{x})^\dagger \phi_\mu^{(0)}(\bm{x})
	\Bigr]\,.
	\label{baryon_density_element}
\end{eqnarray}

To examine the shell structure of the quarks, we evaluate the radial 
density for the $i$th valence quark $\rho^{(i)}(r)$ in which 
the angular degrees of freedom are integrated out, via, 
\begin{eqnarray}
\rho^{(i)}(r)=\int d\varphi\int\sin\theta d\theta~b^{(i)}_{\rm val}
(r,\theta,\varphi)
\end{eqnarray}
with the baryon number 
\begin{eqnarray}
B=\sum_i \int dr r^2\rho^{(i)}(r)\,.
\end{eqnarray}

The profile functions for $B=3-9, 17$ are plotted in Fig.~\ref{fig:profile}. 
In Table~\ref{tab:energy} are the results for the valence quark levels 
as well as the vacuum sea contributions. The valence quark spectra show 
various degenerate patterns depending on the background configuration. 
The results of the ratio of the mass $E_{\rm static}$ to $B\times E^{B=1}_{\rm static}$ 
and comparison to those of the Skyrme model are also in Table.~\ref{tab:energy}. 
They are qualitatively in agreement.   
Using Eq.~(\ref{baryon_density_element}), we estimated the baryon number density
(see Fig.~\ref{fig:bdensity}). As expected the density inherits the same symmetry as the 
corresponding skyrmion.

The valence quark spectra for various $B$ are 
shown in Fig.~\ref{fig:spectrum}. 
These results strongly suggest the existence 
of shell structure for the valence quarks. 
The spectra show (i) four fold degeneracy of the ground state 
labeled by ${\mathcal G}$ and various degenerate pattern for excited levels 
labeled by ${\mathcal A}_1, {\mathcal A}_2,\cdots$, 
(ii) a large energy gap between the ground state ${\mathcal G}$ and 
the first excited level ${\mathcal A}_1$. 
Small dispersions of the spectra are observed in the results. 
In some cases they are caused by the finite size effect of the basis 
(ex. $B=4$). Growing the size $r_{\rm max}$ and increasing the number of the basis, 
more accurate degeneracy will be attained.

In Fig.~\ref{fig:bdrs} are the results of $\rho^{(i)}(r)$ for $B=3-9,17$.
The behaviour of the density near the origin confirms the existence of three shells 
(${\mathcal G},{\mathcal A}_1,{\mathcal A}_2$).
${\mathcal G}$ behaves like ``$S$-wave'' and others like ``$P$-,$D$-wave''
in a hydrogen-like atom.
However most of the densities are nearly on the same surface and very small 
(not zero) near the origin. 
The plateau in the density observed at the center of the nucleus~\cite{bohr} 
can not be attained in our solutions. Therefore one may need to employ the multi-shell 
ansatz~\cite{manton00} even in the case of light nuclei.

The solutions that we obtained here are totally classical one. To reach
the physical observations, one need to take into account the quantum corrections.
Especially, the absolute mass tends to be much higher than the real nucleon mass due 
to a lack of the Casimir effects-- the loop corrections of the order of $O(N_c^{0})$. For $B=1$, 
some attempts for this subject have been done in the Skyrme model~\cite{moussallam91,moussallam93,holzwarth} 
and in the hyblid quark soliton model~\cite{weigel95}. In both cases the calculated 
values properly reduce the large classical energy to the physical nucleon mass. For $B>1$, 
because of the lower symmetry of the chiral fields, the estimations
of these effects are tedious task. Only one work for the $B=2$ chiral soliton 
is reported in Ref.~\cite{scholtz} in which the configuration is restricted to $SO(3)$. 
Since the quantum corrections strongly depend on $B$ it is difficult to extract any information 
of the binding energy from our classical solutions.  

\subsection{Classical Dynamics of Slowly Moving $B=2$ Solitons}

So far our concern has been restricted to static solutions, we shall now 
extend our scope to cover dynamics. Once a multi-soliton solution is obtained, 
one can study the low-energy dynamics 
of the multi-solitons following the geodesic approximation~\cite{manton82}. 
The existence of multi-soliton solutions means that there is no net force 
between the solitons. Manton showed that the multi-BPS monopole solution 
is a consequence of the cancellation of the repulsive force by the exchange 
of gauge bosons and the attractive force by exchange of Higgs bosons~\cite{manton77}. 
The cancellation between the forces can be attained when the solution saturates 
the Bogomol'nyi bound~\cite{bogomol'nyi}. For the Skyrme model, one finds 
\begin{eqnarray}
	E \ge \frac{12\pi^{2}f_{\pi}}{e}B \label{}
\end{eqnarray}
where $e$ is a dimensionless constant which can be fixed experimentally. 
It is known that skyrmions do not saturate this bound and have slightly 
higher energies. 

If the solitons which are far apart initially are given some impact parameter 
and start to move slowly towards each other, they will trace a path close 
to the valley of the potential energy without climbing the wall of the 
potential barrier. The valley is just the parameter space of the soliton solution 
which is called a moduli space. If the solitons move slowly, or adiabatically, 
it will be a good approximation to consider their dynamics only on the moduli 
space (truncation). Then the time evolution of the solitons is approximately 
the geodesics on the moduli space. 

In the $B=2$ rational map ansatz, by imposing axisymmetry 
we get 
\begin{eqnarray}
R_2^{*}=\frac{z^2-a}{-az^2+1}
\label{rmab2}
\end{eqnarray}
where $a$ is a parameter of a geodesic in the moduli space with $-1\le a \le 1$. 
For $a=0$, one recovers the minimal energy toroidal configuration. 
Letting the parameter $a$ be time-dependent, we can examine the adiabatic time evolution 
of initially far-apart two solitons, that is, 
\begin{eqnarray}
	U(r,z;a)=U(r,z;a(t)). \label{}
\end{eqnarray}
The numerical solutions for various values of $a$ are shown in Fig.~\ref{fig:bdensityb2}~\cite{tanaka04}. 
It is interesting that the famous $90^{{\rm o}}$ scattering is observed as in the BPS 
monopole~\cite{gibbons86} and $CP(1)$ solitons~\cite{leese90}. 
However, as can be seen in the energy level shown in Fig.~\ref{fig:energyb2}, the solutions  
have higher energies for larger $a$, which means the moduli motion is not adiabatic since energy costs 
to separate the two solitons. This is due to the fact that the chiral solitons does not saturate 
the so called Bogomol'nyi bound. Thus if the two solitons are initially far apart, they will 
coalesce to form a bound state, forming a torus shape but no scattering. 

\section{\label{sec:level4}Symmetry and the Degeneracy of the Quarks}
The bunch of valence spectra due to the potential with discrete symmetries 
has been observed in the study of heavier nuclear systems. In Ref.~\cite{dudek}, 
the valence spectra are highly degenerate because the deformation 
of the spherically symmetric shell produces large shell gaps.  
Thus the nuclei can be considered to be more stable than the spherical one. 
As discussed in Ref.~\cite{manton98}, the group theory should predict 
the level structure of pion fluctuations. However, our problem is 
more complicated due to the presence of quarks. 
Before discussing it in detail, let us show how the shell deformation 
is related to the degeneracy of the spectrum. 

In general, if an eigenequation given by 
\begin{eqnarray}
H\psi_\mu=E_\mu\psi_\mu
\end{eqnarray}
is invariant under a symmetric operation $\hat{R}$ $\in$ $\hat{g}$, 
the equation transforms as
\begin{eqnarray}
\hat{R}H\psi_\mu=H(\hat{R}\psi_\mu)=E_\mu\psi_\mu\,.
\end{eqnarray}
Therefore the states \{$\psi_\mu, \hat{R}\psi_\mu$\} are degenerate in energy with $E_\mu$. 
The set of $d_\mu$ eigenfunctions \{$\psi_i^{(\mu)}$\}$(i=1,\dots,d_\mu)$ belonging to a given 
eigenvalue $E_\mu$ will provide the 
basis for an irreducible representation of the group $\hat{g}$ of the hamiltonian
\cite{hamermesh}:
\begin{eqnarray}
\hat{R}\psi_j^{(\mu)}=\sum_i\psi_i^{(\mu)} D^{(\mu)}_{ij}(\hat{R})\,.
\end{eqnarray}

The Dirac equation 
\begin{eqnarray}
(i\gamma^\mu \partial_\mu-M)\psi(x)=0
\label{dirac}
\end{eqnarray} 
is a wave equation for fermions with Lorentz-covariance, i.e. 
its form has to be invariant under a transition from one inertial system to 
another one. Let the coordinates of the event be $x^{\mu}$ for an observer A and 
$x^{'\mu}$ for an observer B. Both coordinates are connected by the Lorentz transformation
\begin{eqnarray}
x'=a x~~~{\rm or}~~~ (x')^\nu=a^\nu_\mu x^\mu\,.    
\label{ltrans}
\end{eqnarray}
The equation are invariant under this transformation as 
\begin{eqnarray}
(i\gamma^\mu \partial_\mu-M)\psi(x)=0~~\Leftrightarrow~~(i\gamma'^\mu \partial_\mu-M)\psi'(x')=0\,.
\label{dirac_cov}
\end{eqnarray} 
The Dirac fields are hence transformed via
\begin{eqnarray}
\psi'(x')=\psi'(\hat{a}x)\equiv \hat{S}(\hat{a})\psi(x)=\hat{S}(\hat{a})\psi(\hat{a}^{-1}x')
\label{wf_ltrans}
\end{eqnarray}
and also
\begin{eqnarray}
&&\psi(x)=\hat{S}^{-1}(\hat{a})\psi'(x')=\hat{S}^{-1}(\hat{a})\psi'(\hat{a}x)\,,\nonumber \\
&&\psi'(x')=\hat{S}(\hat{a})\psi'(\hat{a}^{-1}x')~~\Rightarrow~~\psi(x)=\hat{S}(\hat{a}^{-1})\psi'(\hat{a}x)\,.\nonumber   
\end{eqnarray}
One can thus obtain the relation for the transformation operator $\hat{S}(a)$ as 
\begin{eqnarray}
\hat{S}(\hat{a}^{-1})=\hat{S}^{-1}(\hat{a})\, ,
\end{eqnarray}
which reads to  
\begin{eqnarray}
(\hat{S}(\hat{a})i\gamma^\mu \partial_\mu \hat{S}^{-1}(\hat{a})-M)\psi'(x')=0\,.
\label{eq_ltrans}
\end{eqnarray}
In Eq.~(\ref{ltrans}), performing transformation to the coordinates of the 
system B, 
\begin{eqnarray}
\frac{\partial}{\partial x^\mu}=\frac{\partial x'^{\nu}}{\partial x^\mu} \frac{\partial}{\partial x'^{\nu}}
=\hat{a}^{\nu}_{\mu}\frac{\partial}{\partial x'^{\nu}}
\end{eqnarray} 
one gets  
\begin{eqnarray}
(\hat{S}(\hat{a})i\gamma^\mu \hat{S}^{-1}(\hat{a})\hat{a}^{\nu}_{\mu}\partial'_\nu-M)\psi'(x')=0\,.
\label{eq_ltrans2}
\end{eqnarray}
From the covariance of the Dirac equation in Eq.~(\ref{dirac_cov}), $\hat{S}(\hat{a})$ must have the 
following property
\begin{eqnarray}
\hat{a}^{\nu}_{\mu}\gamma^\mu=\hat{S}^{-1}(\hat{a})\gamma^\nu \hat{S}(\hat{a})\,.
\label{gma_ltans}
\end{eqnarray}

Let us consider the transformation law for the equation including the Skyrme chiral fields
\begin{eqnarray}
(i\gamma^\mu \partial_\mu-MU^{\gamma_5}(x))\psi(x)=0\,.
\end{eqnarray} 
If the chiral fields have a point group symmetry $\hat{G}$ such as 
\begin{eqnarray}
U(x')=\hat{G}(\hat{a})U(x)\hat{G}(\hat{a})^\dagger\,,~~(\hat{G}(\hat{a}) \in SU(2)_I)\, ,
\end{eqnarray} 
the Dirac equation is invariant under the Lorentz transformation 
\begin{eqnarray}
(x')^\nu=a^\nu_\mu x^\mu~~
{\rm or}~~x'=\hat{a} x 
\end{eqnarray}
with
\begin{eqnarray}
x'=
\left(\begin{array}{c} 
	~t \\
	~\bm{x}' 
	\end{array}
	\right)\,,
~~\hat{a}=
\left(\begin{array}{cc} 
	~1&~0~ \\
	~0&~\hat{\bm{a}} 
	\end{array}
	\right)  \,,
~~x=
\left(\begin{array}{c} 
	~t \\
	~\bm{x}
	\end{array}
	\right)\,,
\end{eqnarray}
accompanying a corresponding iso-rotation
\begin{eqnarray}
(i\gamma^\mu \partial_\mu-MU^{\gamma_5}(x))\psi(x)=0 
\Leftrightarrow (i\gamma^\nu \partial'_\nu-MU^{\gamma_5}(x'))\psi'(x')=0\,,
\end{eqnarray}
with 
\begin{eqnarray}
\psi'(x')=\hat{K}(\hat{a})\psi(x)\equiv(\hat{S}(\hat{a})\times \hat{G}(\hat{a}))\psi(x)\,.
\end{eqnarray}
The operator $\hat{R}$ corresponding to this rotation is thus defined by
\begin{eqnarray}
\psi'(x)\equiv_{def}\hat{R}\psi(x)=\hat{K}(\hat{a})\psi(\hat{a}^{-1}x)\,.
\label{rotation_R}
\end{eqnarray}
It can be shown that the operator $\hat{R}$ commutes with the hamiltonian using following commutations :
\begin{eqnarray}
&&U^{\gamma_5}(x)\psi(x)=\hat{G}^{-1}U^{\gamma_5}(x')\hat{G}\cdot \hat{R}^{-1}\psi'(x') \nonumber \\
&&\hspace{2cm}=\hat{G}^{-1}\hat{S}^{-1}U^{\gamma_5}(x')\psi'(x')\,, \nonumber \\
&&\hat{R}\psi(x')=\hat{K}\psi(x)~~\to~~\psi(x)=\hat{K}^{-1}\hat{R}\psi(x')\,, \nonumber \\
&&\Rightarrow \hat{K}\hat{K}^{-1}\hat{R}U^{\gamma_5}(x')\psi(x')=U^{\gamma_5}(x')\hat{R}\psi(x')\,, \nonumber \\
&&\hspace{2cm}\therefore~~[\hat{R},U^{\gamma_5}]=0\,, \\
&&\nonumber \\
&&i\gamma^0\gamma^k\partial_k \psi(x)=i\gamma^0\gamma^k a_k^l \partial'_l \psi(x) \nonumber \\
&&\hspace{2cm}=i\hat{S}^{-1}\gamma^0\gamma^l \hat{S} \partial'_l \psi(x) \nonumber \\
&&\hspace{2cm}=i\hat{S}^{-1}\gamma^0\gamma^l \hat{S} \partial'_l \hat{K}^{-1}\psi'(x') \nonumber \\
&&\hspace{2cm}=i\hat{G}^{-1}\hat{S}^{-1}\gamma^0\gamma^l\hat{S} \partial'_l \psi'(x') \nonumber \\
&&\Rightarrow \hat{K} i\gamma^0\gamma^k \partial_k \psi(x)=
i\gamma^0\gamma^k \partial'_k \psi'(x') \nonumber \\
&&\hspace{2cm}\therefore~~[\hat{R},i\gamma^0\gamma^k \partial_k]=0\,.
\end{eqnarray}
which reads $[\hat{R},H]=0$. 

Generally speaking, if eigenequation 
\begin{eqnarray}
H\phi_k=\epsilon_k\phi_k
\end{eqnarray}
is invariant in the group $G$, in terms of the symmetric operation $\hat{R}$ ($\in$ $G$) 
the equation becomes
\begin{eqnarray}
\hat{R}H\phi_k=H(\hat{R}\phi_k)=\epsilon_k\phi_k\,.
\end{eqnarray}
Then, $\phi_k$ and $\hat{R}\phi_k$ are degenerate in energy. 
Thus, constructing $\hat{R}$ for each symmetry of the hamiltonian, one should be able to 
deduce the degeneracy structure of the spectra occurring in the valence level.  

As an example, let us examine a rotational operation for the hedgehog ansatz 
given in Eq.~(\ref{chiral_fields_hedgehog}). The chiral fields exhibit  
spherical symmetry and its infinitesimal spatial rotation is of the form
\begin{eqnarray}
x'^\nu=a^\nu_\mu x^\mu~~\to~~x'^\nu=x^\nu+\epsilon^\nu_\mu x^\mu\,.
\end{eqnarray}
Introducing the small angle $\bm{\epsilon}$ as 
\begin{eqnarray}
\epsilon^2_3=-\epsilon^3_2=\epsilon_1\,,~~
\epsilon^3_1=-\epsilon^1_3=\epsilon_2\,,~~
\epsilon^1_2=-\epsilon^2_1=\epsilon_3\,,~~
\end{eqnarray}
the Dirac field transforms as   
\begin{eqnarray}
\psi(a^{-1}x)&=&\psi(x^\nu-\epsilon^\nu_\mu x^\mu) \nonumber \\
&=&\psi(x^\nu)+\frac{\partial\psi}{\partial x^\nu}(-\epsilon^\nu_\mu) \nonumber \\
&=&\psi(x^\nu)+\bm{\epsilon}\cdot(\bm{x}\times\nabla)\psi(x^\nu) \nonumber \\
&=&(1+i(\bm{x}\times\bm{p})\cdot\bm{\epsilon})\psi(x^\nu)=\exp[i\bm{l}\cdot\bm{\epsilon}]\psi(x^\nu)\,,
\end{eqnarray}
where $\bm{l}=\bm{x}\times\bm{p}$ is orbital angular momentum operator. 
An infinitesimal transformation of the Lorentz and isorotation operators are given by 
\begin{eqnarray}
&&\hat{S}=1-\frac{i}{4}\sigma^{\mu\nu}\epsilon_{\mu\nu}=1+\frac{i}{2}\bm{\Sigma}\cdot\bm{\epsilon}
\,,~~\bm{\Sigma}\equiv
\left(\begin{array}{cc} 
	\bm{\sigma}&0  \\
	0&\bm{\sigma}  \\
	\end{array}
	\right)\,,
\nonumber \\
&&\hat{I}=1+\frac{i}{2}\bm{\tau}\cdot\bm{\epsilon}\,.
\end{eqnarray}
Thus the transformation of the Dirac field under the Lorentz and isorotation is given by 
(see Eq.~(\ref{rotation_R}))
\begin{eqnarray}
\psi'(x)=\hat{R}\psi(x)&=&\hat{S}\times \hat{I}\psi(a^{-1}x) \nonumber \\
&=&(1+\frac{i}{2}\bm{\Sigma}\cdot\bm{\epsilon})(1+\frac{i}{2}\bm{\tau}\cdot\bm{\epsilon})
(1+i(\bm{x}\times\bm{p})\cdot\bm{\epsilon})\psi(x) \nonumber \\
&\sim&(1+i(\frac{1}{2}\bm{\Sigma}+\frac{1}{2}\bm{\tau}+\bm{x}\times\bm{p})\cdot\bm{\epsilon})\psi(x)
=(1+i\bm{K}\cdot\bm{\epsilon})\psi(x)\,.
\end{eqnarray}
This $\bm{K}$ is a grandspin operator and a good quantum number for the hedgehog
hamiltonian. 

The point group symmetry has a discrete operation so that it is not possible to perform
these infinitesimal transformation analysis. Nevertheless, one can see that 
the rotation in Eq.~(\ref{rotation_R}) produces the degenerate eigenstates. 
Let us show briefly the derivation of the rotation operator $\hat{R}$ for $B=3$ tetrahedron and 
the transformation law for the numerical basis constructed in Eq.~(\ref{kahana_ripka}). 
The $B=3$ tetrahedral soliton is characterized by two symmetry operations~\cite{manton98}:
$Z_2\times Z_2$ and $T_d$. 
Specifically, $Z_2\times Z_2$ are characterized by a following M\"obius transformation:
\begin{eqnarray}
z\to -\frac{1}{z}~~\Leftrightarrow~~R(z)\to -\frac{1}{R(z)}\equiv R'(z)
\end{eqnarray}
resultantly, 
\begin{eqnarray}
&&\hat{\bm{n}}_R(z')=(n_1,n_2,n_3) \nonumber \\
&&\to\hat{\bm{n}}'_R(z)=\frac{1}{1+|\it{R}(z)|^2}\left(\rm{2Re}[\it{R}'(z)],
\rm{2Im}[\it{R}'(z)],\rm{1}-|\it{R}'(z)|^2\right) \nonumber \\
&&\hspace{1.6cm}=\frac{1}{1+1/|\it{R}|^2}\left(\rm{2Re}[-1/{\it R}],\rm{2Im}[-1/{\it R}],\rm{1}-|1/\it{R}|^2\right)
\nonumber \\
&&\hspace{1.6cm}=(-n_1,n_2,-n_3)\,.
\end{eqnarray}
The transformation operator $\hat{g}\equiv \exp[-i\frac{\pi}{2}\tau_2]$ ensures
\begin{eqnarray}
\hat{g}(\bm{\tau}\cdot \bm{n}_R)\hat{g}^{\dagger}&=&(-i\tau_2)
(\tau_1n_1+\tau_2n_2+\tau_3n_3)(i\tau_2) \nonumber \\
&=&-n_1 \tau_1+n_2 \tau_2-n_3 \tau_3\equiv \bm{\tau}\cdot \bm{n'}_R\,.
\end{eqnarray}
and transforms the chiral field as
\begin{eqnarray}
U(\bm{x}')=\hat{g}U(\bm{x})\hat{g}^{\dagger}\,.
\label{htrans}
\end{eqnarray}
$T_d$ transforms the $z$ and $R(z)$ as  
\begin{eqnarray}
z\to \frac{iz+1}{-iz+1}\equiv z'~~\Leftrightarrow~~R(z) \to \frac{iR+1}{-iR+1}\equiv R'(z)\,,
\end{eqnarray}
and hence 
\begin{eqnarray}
&&n_1+in_2 \nonumber \\
&&\to n'_1+in'_2=\frac{2R'}{1+|R'|^2}
=\frac{1+i(R+\bar{R})-|R|^2}{1+|R|^2}=n_3+in_1\,, \nonumber \\
&&n_1-in_2 \nonumber \\
&&\to n'_1-in'_2=\frac{2\bar{R'}}{1+|R'|^2}
=\frac{1-i(R+\bar{R})-|R|^2}{1+|R|^2}=n_3-in_1\,, \nonumber \\
&&n_3 \to n'_3=\frac{1-|R'|^2}{1+|R'|^2}
=\frac{-i(R-\bar{R})}{1+|R|^2}=n_2\,,
\end{eqnarray}
which yields  
\begin{eqnarray}
n'_1\tau_1+n'_2\tau_2+n'_3\tau_3&=&\frac{1}{2}(n'_1+in'_2)(\tau_1-i\tau_2)
+\frac{1}{2}(n'_1-in'_2)(\tau_1+i\tau_2)+n'_3\tau_3 \nonumber \\
&=&\frac{1}{2}(n_3+in_1)(\tau_1-i\tau_2)
+\frac{1}{2}(n_3-in_1)(\tau_1+i\tau_2)+n_2\tau_3 \nonumber \\
&=&n_3\tau_1+n_1\tau_2+n_2\tau_3\,.
\end{eqnarray}
The transformation operator 
$\hat{h}\equiv \exp[-i\frac{\pi}{3}\frac{1}{\sqrt{3}}(\tau_1+\tau_2+\tau_3)]=(1-i(\tau_1+\tau_2+\tau_3))/2$
ensures
\begin{eqnarray}
\hat{h}(\bm{\tau}\cdot \bm{n}_R)\hat{h}^{\dagger}&=&\frac{1}{2}(1-i(\tau_1+\tau_2+\tau_3))
(\tau_1n_1+\tau_2n_2+\tau_3n_3)\frac{1}{2}(1+i(\tau_1+\tau_2+\tau_3)) \nonumber \\
&=&n_3\tau_1+n_1\tau_2+n_2\tau_3\equiv \bm{\tau}\cdot \bm{n'}_R\, ,
\end{eqnarray}
and transforms the chiral field as
\begin{eqnarray}
U(\bm{x}')=\hat{h}U(\bm{x})\hat{h}^{\dagger}\,.
\label{gtrans}
\end{eqnarray}

The Lorentz transformation operator $\hat{S}$ and the operators for 
the chiral fields $\{\hat{g},\hat{h}\}$ corresponding to the symmetric operations 
$(x')^\nu=a^\nu_\mu x^\mu$ are given by
\begin{eqnarray}
&&Z_2\times Z_2: \nonumber \\
&&\hspace{1cm}(a_g)^\nu_\mu=
\left(\begin{array}{cccc} 
	~1&~0&~0&~0 \\
	~0&-1&~0&~0 \\
	~0&~0&~1&~0 \\
	~0&~0&~0&-1 \\
	\end{array}
	\right)~~\Rightarrow~~
\hat{S}_g=i\gamma^0\gamma^5\gamma^2\,,  \\
&&T_d: \nonumber \\
&&\hspace{1cm}(a_h)^\nu_\mu=
\left(\begin{array}{cccc} 
	~1&~0&~0&~0 \\
	~0&~0&~1&~0 \\
	~0&~0&~0&~1 \\
	~0&~1&~0&~0 \\
	\end{array}
	\right)  
\Rightarrow
\hat{S}_h=\exp[i\frac{\pi}{3}\frac{1}{\sqrt{3}}(\sigma_{23}+\sigma_{31}+\sigma_{12})]\,.
\nonumber \\
\end{eqnarray}
The $\hat{R}$ is defined by the direct product of these rotation operators together
with the inverse spatial rotation for the spinor such as
\begin{eqnarray}
&&\hat{R}_g\psi(x)\equiv (\hat{S}_g\times \hat{g})\psi(\hat{a}_g^{-1}x)\,, \\
&&\hat{R}_h\psi(x)\equiv (\hat{S}_h\times \hat{h})\psi(\hat{a}_h^{-1}x)\,.
\end{eqnarray} 
Applying these operators to the Kahana-Ripka basis $\phi\equiv\{u,v\}$ (for detail, 
see Appendix B) we finally obtain the following transformation laws:
\begin{eqnarray}
&&\hat{R}_g\phi_{KM}=(-1)^{K-M}\phi_{K-M}\,. 
\label{basetransz}
\\
&&\hat{R}_h\phi_{00}=\phi_{00}\,. 
\label{basetrans0}
\\
&&\hat{R}_h\left(
\begin{array}{c}
\phi_a \\\phi_b \\\phi_c
\end{array}
\right)=
\left(
\begin{array}{ccc}
~0&-1&~0 \\
~0&~0&-1 \\
~1&~0&~0  \\
\end{array}
\right)
\left(
\begin{array}{c}
\phi_a \\\phi_b \\\phi_c
\end{array}
\right)\,, \\
&&\phi_a\equiv\frac{1}{\sqrt{2}}(\phi_{11}+\phi_{1-1})\,,~~
\phi_b\equiv i\phi_{10}\,,~~
\phi_c\equiv \frac{i}{\sqrt{2}}(\phi_{11}-\phi_{1-1})\,.
\label{basetrans1}
\nonumber \\
&&\hat{R}_h\left(
\begin{array}{c}
\phi_\xi \\\phi_\eta \\\phi_\zeta \\\phi_u \\\phi_v
\end{array}
\right)=
\left(
\begin{array}{ccc|cc}
~0&-1&~0~& 0& 0\\
~0&~0&~1~& 0& 0\\
-1&~0&~0~& 0& 0\\ \hline
 0& 0& 0&-\frac{1}{2}&~\frac{\sqrt{3}}{2}  \\
 0& 0& 0&-\frac{\sqrt{3}}{2} & -\frac{1}{2}\\
\end{array}
\right)
\left(
\begin{array}{c}
\phi_\xi \\\phi_\eta \\\phi_\zeta \\\phi_u \\\phi_v
\end{array}
\right)\,, \\
&&\phi_\xi=\frac{i}{\sqrt{2}}(\phi_{21}+\phi_{2-1})\,,~~
\phi_\eta=\frac{1}{\sqrt{2}}(\phi_{21}-\phi_{2-1})\,, \nonumber \\
&&\phi_\zeta=\frac{i}{\sqrt{2}}(\phi_{22}-\phi_{2-2})\,,~~
\phi_u=\phi_{20}\,,~~
\phi_v=\frac{1}{\sqrt{2}}(\phi_{22}+\phi_{2-2})\,, 
\label{basetrans2}
\nonumber 
\end{eqnarray}
confirming that the $B=3$ tetrahedron can exhibit triply degenerate spectra.

The numerical computation indicates that the winding number strongly couple 
the elements with different $K$ and correlated valence spectra occur as a result.  
As can be seen from the operator $K$ with $B=2$, the degeneracy of the valence 
spectra are explained in terms of the shape deformation (symmetry) as well as the winding number  
of the chiral fields~\cite{sawado98}. The four-fold degeneracy of the lowest states 
may be ascribed to the chiral symmetry $SU(2)_L\times SU(2)_R$ of the hamiltonian. 
The degenerate structure for $B\ge 3$ will be well understood if symmetric operators 
of the hamiltonian which consist of the angular momentum, spin, isospin and winding 
number, are explicitly constructed. 

\section{\label{sec:level5}Zero Mode Quantization}
The solitons that we obtained in the previous section are 
classical objects and therefore must be 
quantized to assign definite spin and isospin to them.  
Quantization of the solitons can be performed semiclassically 
for their rotational zero modes. 
For the hedgehog soliton, because of its topological structure, 
a rotation in isospin space is followed by a simultaneous spatial 
rotation. For the axially symmetric soliton, 
there are five rotational zero modes by rotations of  
iso-degrees of freedom and spatial rotations. 
\subsection{$SU(2)$ sector}
Let us introduce the dynamically rotated chiral fields around the classical fields $U_S$
\cite{braaten88}:
\begin{equation}
	U(\bm{x},t)=A(t)U_S(\bm{x}')A(t)^{\dagger},~~{{x}^{i}}'  = 
	{\Xi}^{i}_j[B(t)]x^{j}
	\label{eq:35}
\end{equation}
where 
\begin{equation}
{\Xi}^{i}_j[B(t)] =\frac{1}{2}{\rm Tr}[{\sigma}^{i}B(t){\sigma}_{j}B(t)^{\dagger}] \,,
\end{equation}
and $A(t)$ and $B(t)$ are time-dependent $SU(2)$ matrices generating 
an iso-rotation and a spatial rotation respectively. 
By transforming the rotating frame of reference, the Dirac operator 
with Eq.~(\ref{eq:35}) can be written as 
\begin{eqnarray}
	\tilde{iD}&=& i \gdhi - M U^{\gamma_5}(\bm{x},t) \nonumber \\
	&=& A(t)S(t)^{\dagger}{\gamma}^0 [i {\partial}_t + i 
	\tilde{{\gamma}^{0}}\tilde{{\gamma}^{k}} {\partial_k}'-MU_S^{\gamma_5}(\bm{x}') +
	iA^{\dagger} \dot{A}+iS^{\dagger} \dot{S}]S(t)A(t)^{\dagger}
	\label{eq:37} \nonumber \\
\end{eqnarray}
where 
\begin{equation}
	\tilde{{\gamma}^{\mu}}= {\Lambda}^{\mu}_{\nu} S {\gamma}^{\nu} S^{\dagger} 
	=  \left( 
	\begin{array}{c}
	{\gamma}^{0} \\
	{\gamma}^{k} + (\bm{r}' \times \dot{\bm{\theta}})^{k}{\gamma}^{0} 
	\end{array}\right) \,,
	\label{eq:38}
\end{equation}
and $S(t)$ is the rotation operator for the Dirac field 
and $\bm{\theta}$ is an angle of the spatial rotation. 
Note that the gamma matrices $\tilde{{\gamma}^{\mu}}$ explicitly depend 
on the coordinates and do not transform as a contravariant 
vector~\cite{koepf89}. 
Substituting Eq.~(\ref{eq:38}) into Eq.~(\ref{eq:37}), one obtains
\begin{eqnarray}
	\tilde{iD}= A(t)S(t)^{\dagger}{\gamma}^0 [i {\partial}_t -
	H(U_S^{{\gamma}_{5}}) + {\Omega}_{A} + {\Omega}_{B} ]S(t)A(t)^{\dagger} 
\end{eqnarray}
where 
\begin{eqnarray}
	\Omega_{A}=i{A^{\dagger}} \dot{A}=\frac{1}{2} \Omega^{a}_{A} \tau_a\,,\,\, 
	\Omega_{B}=i{S^{\dagger}} \dot{S} + (\bm{r} \times \bm{p}) \cdot 
	\dot{\bm{\theta}}= \Omega^{a}_{B} J_b
\end{eqnarray}
with $J_{a}=1/2 {\epsilon}_{abc} {\gamma}^{b} {\gamma}^{c} - i( \bm{r} \times \nabla )_{a}$. 
$\Omega_{A}$ and $\Omega_{B}$ are the angular velocity operators for an isorotation 
and for a spatial rotation respectively.   
Under these operations the effective action can be written by 
\begin{eqnarray}
	S_{\rm eff}(U) &=& S_{\rm eff}(U_S) \nonumber \\
	&-&iN_c\Sp\left[
	\log \bigl(i {\partial}_t-H(U^{{\gamma}_{5}}_{S}) 
	+ {\Omega}_{A} + {\Omega}_{B}\bigr)\right]
	-\Sp\left[\log (i {\partial}_t 
	-H(U^{{\gamma}_{5}}_{S})\right] \nonumber \\
	&\to& S_{\rm eff}(U_S)
	-\frac{1}{2}iN_c\Sp \log dd^\dagger
	+\frac{1}{2}iN_c\Sp \log d_0{d_0}^\dagger\,.
\end{eqnarray}
With the proper-time regularization, it reads
\begin{eqnarray}
S^{\rm reg}_{\rm eff}(U)=S^{\rm reg}_{\rm eff}(U_S)
-\frac{N_c}{2}\int \frac{d\omega}{2\pi}\int^\infty_{1/\Lambda^2} \frac{d\tau}{\tau}
\Sp[e^{-dd^\dagger\tau}-e^{-d_0{d_0}^\dagger\tau}]\,,
\end{eqnarray}
where 
\begin{eqnarray}
&&d=i\omega-H(U_S^{\gamma_5})+\Omega_A+\Omega_B,
d^\dagger=-i\omega-H(U_S^{\gamma_5})-\Omega_A-\Omega_B \, ,\\
&&dd^\dagger=\omega^2+H^2-2i\omega(\Omega_A+\Omega_B)-[H,\Omega_A+\Omega_B]
-(\Omega_A+\Omega_B)^2\,,\\
&&d_0{d_0}^\dagger=\omega^2+H^2\,.
\end{eqnarray}
Assuming that the rotation of the soliton is adiabatic,  
we shall expand the effective action $S_{\rm eff}$ around the classical 
solution $U_{S}$ with respect to the angular momentum velocity 
${\Omega}_{A}$ and ${\Omega}_{B}$ up to second order~\cite{biedenharn85}
\begin{eqnarray}
	S^{\rm reg}_{\rm eff}(U)&=&S^{\rm reg}_{\rm eff}(U_S) \nonumber \\
	&+&\frac{1}{2} \sum_{ab}\int dt \bigl[
	  I^{AA}_{0,ab} {\Omega}_{A}^{a}(t) {\Omega}_{A}^{b}(t) 
  	+ I^{AB}_{0,ab} {\Omega}_{A}^{a}(t) {\Omega}_{B}^{b}(t) \nonumber \\
	&+& I^{BA}_{0,ab} {\Omega}_{B}^{a}(t) {\Omega}_{A}^{b}(t)
	+ I^{BB}_{0,ab} {\Omega}_{B}^{a}(t) {\Omega}_{B}^{b}(t) \bigr]
\end{eqnarray}
where $I_{0,ab}$'s are the vacuum sea contributions to the moments of inertia
defined by 
\begin{eqnarray*}
	&&I^{AA}_{0,ab} = \frac{1}{8}N_c \sum_{n,m}f(E_m,E_n,\Lambda) 
	{\langle n|{\tau}_a | m \rangle} 
	{\langle m|{\tau}_b |n \rangle}\,, \\
	&&I^{AB}_{0,ab} = \frac{1}{4}N_c \sum_{n,m}f(E_m,E_n,\Lambda){\langle n|{\tau}_a | m \rangle}
	{\langle m|{J}_b |n \rangle}\,, \\
	&&I^{BA}_{0,ab} = \frac{1}{4}N_c \sum_{n,m}f(E_m,E_n,\Lambda)
	{\langle n|{J}_a | m \rangle} 
	{\langle m|{\tau}_b |n \rangle}\,, \\
	&&I^{BB}_{0,ab} = \frac{1}{2}N_c \sum_{n,m}f(E_m,E_n,\Lambda)
	{\langle n|{J}_a | m \rangle} 
	{\langle m|{J}_b |n \rangle}  
	\label{eq:inertiao}
\end{eqnarray*}
with the cutoff function $f(E_m,E_n,\Lambda)$ 
\begin{eqnarray}
	f(E_m,E_n,\Lambda)&=&-\frac{2\Lambda}{\sqrt{\pi}}
	\frac{e^{-E_m^2/\Lambda^2}-e^{-E_n^2/\Lambda^2}}{E_m^2-E_n^2} \nonumber \\
	&+&\frac{{\rm sgn}(E_m){\rm erfc}(|E_m|/ \Lambda)
	-{\rm sgn}(E_n){\rm erfc}(|E_n|/ \Lambda)}{E_m-E_n}\,. 
\label{cutoff}
\end{eqnarray}
Similarly, for the valence quark contribution to the moments of inertia, 
we have
\begin{eqnarray}
	&&I^{AA}_{{\rm val},ab} = \frac{1}{2}N_c \sum_{m \neq {\rm val}} 
	\frac{{\langle {\rm val}|{\tau}_a | m \rangle} {\langle m|{\tau}_b |{\rm val} \rangle}}
	{E_m - E_{\rm val}}\,, \nonumber \\
	&&I^{AB}_{{\rm val},ab} = N_c \sum_{m \neq {\rm val}} 
	\frac{{\langle {\rm val}|{\tau}_a | m \rangle}{\langle m|{J}_b | {\rm val} \rangle}}
	{E_m - E_{\rm val}}\,, \nonumber \\
	&&I^{BA}_{{\rm val},ab} = N_c \sum_{m \neq {\rm val}} 
	\frac{{\langle {\rm val}|{J}_a | m \rangle} {\langle m|{\tau}_b | {\rm val} \rangle}}
	{E_m - E_{\rm val}}\,, \nonumber \\
	&&I^{BB}_{{\rm val},ab} = 2N_c \sum_{m \neq {\rm val}} 
	\frac{{\langle {\rm val}|{J}_a | m \rangle} {\langle m|{J}_b | {\rm val} \rangle}}
	{E_m - E_{\rm val}}\,.
	\label{eq:inertiav}
\end{eqnarray}
The total moments of inertia are then given by the sum of the vacuum 
and valence as $I^{AA}_{ab} = I^{AA}_{{\rm val},ab} + I^{AA}_{0,ab}$.

Finally, the effective lagrangian is obtained as 
\begin{eqnarray}
L=-E_{\rm static}+\frac{1}{2}I^{AA}_{ab} \Omega^a_A\Omega^b_A
+I^{AB}_{ab} \Omega^a_A\Omega^b_B+\frac{1}{2}I^{BB}_{ab} \Omega^a_B\Omega^b_B\,.
\end{eqnarray}

Theoretically, these moments of inertia can be computed using the 
eigenstates of Eq.~(\ref{eigenequation}). However, due to the difference 
of the boundary conditions between the initial and final states of 
the matrix element, the moments of inertia acquire nonzero values 
with vanishing pion fields. 
To overcome this problem, we make the following replacement~\cite{goeke91}:
\begin{eqnarray}
	{\langle n |{J}_a | m \rangle} &\rightarrow& {\langle n |
	[H(U^{\gamma_5}_S),{J}_a] | m \rangle}/(E_n - E_m) \nonumber\\
	&=&{\langle n |[MU^{\gamma_5}_S ,{l}_a] | m \rangle}/(E_n - E_m)
	\label{eq:replacement}
\end{eqnarray}
where ${l}_a=-i(\bm{r} \times \nabla)_a$.  
Unless the hamiltonian explicitly depend on the coordinates, 
the numerator vanishes with vanishing pion fields. 
The spurious contributions to the moment of 
inertia can be removed in this way.

From axial symmetry of the system, following relations are derived
\begin{eqnarray}
	&&\hspace{1.8cm}I_{ij}= 0, ~~i\neq j\,;\,\,	
	I_{11}^{AB} =  I_{22}^{AB} = I_{11}^{BA} =  I_{22}^{BA} = 0\,,\nonumber \\
	&&I_{11}^{AA} =  I_{22}^{AA}, ~~ I_{11}^{BB} =  I_{22}^{BB}\,,\,\,
	I_{33}^{BB} = m^{2} I_{33}^{AA}, ~ I_{33}^{AB} =  I_{33}^{BA} = -m_{{\rm w}} I_{33}^{AA}.
	\label{eq:50}
\end{eqnarray}

The quantization conditions for the collective coordinates, $A(t)$ and $B(t)$, 
define a body-fixed isospin operator $\bm{K}$ and a body-fixed angular 
momentum operator $\bm{L}$ as 
\begin{eqnarray}
	&&I^{AA}_{ab} {\Omega}^{b}_{A} + I^{AB}_{ab} {\Omega}^{b}_{B} \rightarrow -
	\tr \bigg( A \frac{ {\tau}_a }{2} \frac{\partial}{\partial A}  \biggr) 
	\equiv -K_a\,, \\
	&&I^{BA}_{ab} {\Omega}^{b}_{A} + I^{BB}_{ab} {\Omega}^{b}_{B} \rightarrow 
	\tr \bigg( \frac{ {\sigma}_a }{2} B \frac{\partial}{\partial B}  \biggr) 
	\equiv -L_a.
	\label{eq:48}
\end{eqnarray}
These are related to the usual coordinate-fixed isospin 
operator $I_a$ and coordinate-fixed angular momentum $J_a$ operator by 
transformations,
\begin{equation}
	I_{a}=- \Xi^{b}_{a}[A(t)]K_b,~~~J_{a}=- \Xi^{b}_{a}[B(t)]^{T}L_b .
	\label{eq:orthotrans}
\end{equation}    
   
To estimate the quantum energy corrections, let us introduce 
the basis functions of the spin and isospin operators which
were inspired from the cranking method for 
nuclei~\cite{bohr,weigel86},
\begin{eqnarray*}
	&&\langle A,B|{i{i}_{3}{k}_{3},j{j}_{3}{l}_{3}}\rangle
	=\frac{\sqrt{(2i+1)(2j+1)}}{8{\pi}^{2}}
	D^{i~~\ast}_{i_{3}k_{3}}(A)D^{j~~\ast}_{j_{3},-m_{\rm w}k_{3}}(B) 
\end{eqnarray*}  
where $D$ is the Wigner rotation matrix. Then, we find the 
quantized energies of the soliton as 
\begin{eqnarray}
	&&E=E_{\rm static} \nonumber \\
	&&+\frac{1}{2I_{11}^{AA}}i(i+1)
	+\frac{1}{2I_{11}^{BB}}j(j+1) 
	+\frac{1}{2}\left(\frac{1}{I_{33}^{AA}}-\frac{1}{I_{11}^{AA}}
	-\frac{m^2_{\rm w}}{I_{11}^{BB}}\right)k_{3}^{2} \label{eq:51}
\end{eqnarray}
where $i(i+1), j(j+1)$ and $k_3$ are eigenvalues of the Casimir operators 
$\bm{I}^2$ and $\bm{J}^2$, and the operator $\bm{K}_3$, respectively.

In Table~\ref{tab:table3} are the results of our calculation 
of moments of inertia, $I_{11}^{AA}, I_{11}^{BB}$ and $I_{33}^{AA}$, 
with $B=2-5$. It is instructive to compare our results
with the Skyrme model~\cite{braaten88} where $U_{11}=0.0104, 
V_{11}=0.0163$ and $U_{33}=0.00709$ which are correspondingly 
our $I_{11}^{AA}, I_{11}^{BB}$ and $I_{33}^{AA}$. 
They are qualitatively in good agreement.  

\begin{table}
\caption{\label{tab:table3}Moments of inertia~(in MeV$^{-1}$).}
\begin{tabular}{cccccccccc} \hline
~~~$B$& & Valence & Sea & Total& $B$& & Valence & Sea & Total \\
\hline
~~~2&$I^{AA}_{11}$&0.00773& 0.00363 &0.01136  &4&$I^{AA}_{11}$&0.01408& 0.00959 &0.02366\\
~~~ &$I^{BB}_{11}$&0.01141& 0.00464 &0.01605  & &$I^{BB}_{11}$&0.04272& 0.01245 &0.05517\\
~~~ &$I^{AA}_{33}$&0.00429& 0.00125 &0.00554 & &$I^{AA}_{33}$&0.01172& 0.00074 &0.01246\\
\hline
~~~3&$I^{AA}_{11}$&0.01231& 0.00280 &0.01511  &5&$I^{AA}_{11}$&0.02786& 0.00716 &0.03502\\
~~~ &$I^{BB}_{11}$&0.02174& 0.00384 &0.02558  & &$I^{BB}_{11}$&0.12124& 0.01112 &0.13236\\
~~~ &$I^{AA}_{33}$&0.00594& 0.00027 &0.00622 & &$I^{AA}_{33}$&0.01368& 0.00007 &0.01375 \\
\hline
\end{tabular}
\end{table}

\subsection{Finkelstein-Rubinstein constraints}
If a multi-skyrmion describes atomic nuclei upon quantization,
it has to be quantized as a boson or as a fermion whether $B$ is even or odd. 
This requirement is implemented in the form of Finkelstein-Rubinstein 
(FR) constraints \cite{fr}. For highly nonlinear theory enough to possess soliton 
solutions a consideration of continuity reduces to a concept of distinct topological sector.  
Finkelstein and Rubinstein state that an equally primitive concept as continuity is  
between multi-valued quantized systems which can possess state functions double-valued 
under $2\pi$ rotation, and those which cannot.
Both concepts are ascribed to homotopy of the map between the physical space 
and the configuration space. 
Indeed for the $SU(2)$ chiral soliton solution to exist, the physical space must be 
compactified to $S^{3}$ which defines a topological charge characterized by an integer 
\begin{eqnarray}
	\pi_{3}(S^{3})=n\, . \label{}
\end{eqnarray}
The FR constraints arise when the space-time is suitably compactified as 
\begin{eqnarray}
	\pi_{4}(SU(2))=Z_{2}\, , \label{}
\end{eqnarray}
which takes values only $-1$ or $+1$. This allows us to quantize the solitons 
as either a fermion or a boson. 

The FR constraints for the rational map ansatz 
was constructed in Ref.~\cite{irwin} and Ref.~\cite{krusch} and applied to predict 
the ground states of skyrmions up to $B=22$. In this section, we shall 
apply the FR constraints for the rational map ansatz directly to our axially 
symmetric multi-skyrmions and obtain their ground states. 
 
Following the notation in Ref.~\cite{krusch}, let $g$ be a rotation by $\alpha$ 
around $\bm{n}$ followed by an isorotation by $\beta$ around $\bm{N}$. 
Then the FR constraints can be defined by  
\begin{eqnarray}
	\exp(-i\alpha \bm{n}\cdot\bm{J})\exp(-i\beta \bm{N}\cdot\bm{I})
	\psi =  \chi_{FR}(g)\psi \label{}
\end{eqnarray}
where 
\begin{eqnarray}
\chi_{FR}(g)=\left\{
\begin{array}{l}
~1~~~~{\rm if~contractible} \\
-1~~~{\rm otherwise}.
\end{array}\right.
\end{eqnarray}
and, $\bm{J}$ and $\bm{I}$ are space-fixed spin and isospin operators respectively.
$\psi$ is the wave function which transforms under a tensor product 
of rotations and isorotations. In particular, a closed loop is noncontractible 
for odd $B$ and contractible for even $B$, which is consistent with spin statistics. 
Consequently, quantum numbers $I$ and $J$ are half-integers for odd $B$ and 
integers for even $B$. 

In order to construct the ground states for a given baryon number $B$, 
let us define $N(L(\alpha, \beta))$ as a homotopy invariant for a loop $L$
generated by rotations by $\alpha$ and isorotations by $\beta$. 
Then, for the axially symmetric rational map of degree $B$, it is given by Ref.~\cite{krusch}
\begin{eqnarray}
	N(L(\alpha, \beta))=\frac{B}{2\pi}(B\alpha-\beta) \,.\label{n(l)}
\end{eqnarray}
It can be shown that $N$ (mod $2$) determines if the loop is contractible or 
not in the same sense as $B$ (mod $2$). Therefore, $N$ (mod $2$) gives the FR 
constraints for each generator of the symmetry group of the rational map. 

The axially symmetric rational map with degree $B$ is given by 
\begin{eqnarray}
	R(z)=\frac{1}{z^{B}} \, . \label{}
\end{eqnarray}
There are two symmetric generators for this rational map. 
One is a rotation by $\alpha$ followed by an isorotation by 
$\beta = B\alpha$. Substituting it into Eq.~(\ref{n(l)}), one obtains $N(L(\alpha, 
B\alpha))=0$. The FR constraints for this loop is thus given by 
\begin{eqnarray}
	e^{-i\pi(L_{3}-BK_{3})}\psi = \psi \, . \label{j3}
\end{eqnarray} 
where $L_3$, $K_3$ are the third component of the body-fixed angular 
momentum and the isospin operators which are related with 
the space-fixed operators by orthogonal transformations (\ref{eq:orthotrans}). 
The other symmetry is $C_{2}$ with transformation 
\begin{eqnarray}
	z \rightarrow \frac{1}{z}\;, \;\;\; R(z) \rightarrow \frac{1}{R(z)}\,. \label{}
\end{eqnarray}
This corresponds to $\alpha=\beta=\pi$ and hence $N(L(\pi,\pi))=B(B-1)/2$. 
The FR constraints for this loop is 
\begin{eqnarray}
	e^{-i\pi(L_{1}+K_{1})}\psi = (-1)^{B(B-1)/2}\psi \,. \label{j1}
\end{eqnarray}

In the following we construct the ground states consistent with the derived 
FR constraints (\ref{j3}) and (\ref{j1}) for $B=2-5$ with axial symmetry. 
\begin{itemize}
	\item $B=2$ \\
	We find the FR constraints 
\begin{eqnarray}
	&& e^{-i\pi(L_{3}-2K_{3})}\psi = \psi \\ 
	&& e^{-i\pi(L_{1}+K_{1})}\psi = -\psi \,. \label{}
\end{eqnarray} 
This gives the ground state as 
$\left|J,L_{3}\right>\left|I,K_{3}\right>=\left|1,0\right>\left|0,0\right>$. 
This is in agreement with the ground state ${}^1_1$H$^+$~(deuteron).\\
	\item $B=3$ \\
	We find the FR constraints 
\begin{eqnarray}
	&& e^{-i\pi(L_{3}-3K_{3})}\psi = \psi \\
	&&e^{-i\pi(L_{1}+K_{1})}\psi = -\psi  
\end{eqnarray} 
This gives the ground state as 
$\left|\frac{5}{2},\frac{3}{2}\right>\left|\frac{1}{2},\frac{1}{2}\right>$. 
The ground state~(and first ``excited'' state) of $B=3$ have 
$(I,J)=(\frac{1}{2},\frac{1}{2})$~(${}^3_2$He$^+$, ${}^3_1$H$^+$). Then, one
can not identify our soliton with these observed isodoublet nuclei. \\ 
	\item $B=4$ \\
	We find the FR constraints 
\begin{eqnarray}
	&& e^{-i\pi(L_{3}-4K_{3})}\psi = \psi \\ 
	&& e^{-i\pi(L_{1}+K_{1})}\psi = \psi \,. \label{}
\end{eqnarray} 
This gives the ground state as $\left|0,0\right>\left|0,0\right>$. 
The ground state ${}^4_2$He$^+$ has 
$(I,J)=(0,0)$ . Then our soliton can be identified as 
the ``$\alpha$ particle''. \\ 
	\item $B=5$ \\
	We find the FR constraints 
\begin{eqnarray}
	&& e^{-i\pi(L_{3}-5K_{3})}\psi = \psi \\ 
	&& e^{-i\pi(L_{1}+K_{1})}\psi = \psi \,. 
	\label{}
\end{eqnarray} 
This gives the ground state as $\left|J\right>\left|I\right>=\left|
\frac{7}{2},\frac{5}{2}\right>\left|\frac{1}{2},\frac{1}{2}\right>$. 
This is not in agreement with the observed nuclei of $B=5$ ;
$(I,J)=(\frac{1}{2},\frac{3}{2})$~(${}^5_2$He$^+$, ${}^5_3$Li$^+$).
\\ 
	
\end{itemize}
Thus, we conclude that for even $B$, the axially symmetric solitons 
are possible candidates of the ground states of $B$ atomic nuclei 
as is the case of the deuteron and ${}^4_2$He while for odd $B$ 
they emerge only as excited states. 

\begin{table}
	\begin{center}
	\caption{\label{tab:b2}$B=2$, mass spectrum up to $i,j\le 3, k_3\le 1$.}
	\begin{tabular}{cccc} \hline
	~~Classification & ($i,j,k_3$) & Parity & Mass[MeV] \\
	\hline
	$NN(^{3}{S}_{1})$              &($ 0,1,0$)& $+$ &2264\\
	$NN(^{1}{S}_{0})$              &($ 1,0,0$)& $+$ &2290\\
	$N\Delta(^{3}{P}_{2})$         &($ 1,2,1$)& $-$ &2399\\
	$N \Delta (^{5}{S}_{2})$       &($ 1,2,0$)& $+$ &2477\\
	$N \Delta (^{3}{S}_{1})$       &($ 2,1,0$)& $+$ &2528\\
	$ \Delta \Delta (^{7}{S}_{3})$ &($ 0,3,0$)& $+$ &2576\\
	$ \Delta \Delta (^{1}{S}_{0})$ &($ 3,0,0$)& $+$ &2730\\
	$\Delta\Delta(^{5}{P}_{3})$    &($ 2,3,1$)& $-$ &2762\\ \hline
	\end{tabular}
	\end{center}
\end{table}

\begin{table}
	\begin{center}
	\caption{\label{tab:b4}$B=4$, mass spectrum up to $i\le 3,j\le 5,k_3\le 1$.}
	\begin{tabular}{cccc} \hline
	~~Classification & ($i,j,k_3$) & Parity & Mass[MeV] \\
	\hline
	$^{4}N(^{1}{S}_{0})$              &($ 0,0,0$)& $+$ &4753\\
	$^{4}N(^{5}{S}_{2})$              &($ 0,2,0$)& $+$ &4807\\
	$^{2}N~^{2}\Delta(^{7}{P}_{4})$   &($ 0,4,1$)& $-$ &4808\\
	$^{4}N(^{3}{S}_{1})$              &($ 1,1,0$)& $+$ &4813\\
	$^{4}N(^{1}{S}_{0})$              &($ 2,0,0$)& $+$ &4879\\
	$^{3}N~\Delta(^{7}{S}_{3})$       &($ 1,3,0$)& $+$ &4904\\
	$^{4}N(^{5}{S}_{2})$              &($ 2,2,0$)& $+$ &4934{\tiny.2}\\
	$^{2}N~^{2}\Delta(^{7}{S}_{4})$   &($ 0,4,0$)& $+$ &4934{\tiny.3}\\
	$^{2}N~^{2}\Delta(^{9}{P}_{4})$   &($ 2,4,1$)& $-$ &4935\\
	$N~^{3}\Delta(^{9}{P}_{5})$       &($ 1,5,1$)& $-$ &4941\\
	$^{3}N~\Delta(^{3}{S}_{1})$       &($ 3,1,0$)& $+$ &5025\\
	$N~^{3}\Delta(^{9}{S}_{5})$       &($ 1,5,0$)& $+$ &5046\\
	$^{2}N~^{2}\Delta(^{9}{S}_{4})$   &($ 2,4,0$)& $+$ &5061\\
	$^{3}N~\Delta(^{7}{S}_{3})$       &($ 3,3,0$)& $+$ &5115\\
	$N~^{3}\Delta(^{9}{P}_{5})$       &($ 3,5,1$)& $-$ &5152\\
	$N~^{3}\Delta(^{9}{S}_{5})$       &($ 3,5,0$)& $+$ &5278\\ \hline
	\end{tabular}
	\end{center}
\end{table}

The results of the quantized energy of the axially symmetric soliton solutions 
with $B=2, 4$ are shown in Table \ref{tab:b2} and \ref{tab:b4}. 
The study of the Finkelstein-Rubinstein constraints indicates that 
the axially symmetric solution with even $B$ has the same quantum number 
as the physically observed nuclei. Some of the states may be observed in experiments. 
Specifically, in the $B=2$, we obtained the $I=0,J=1$ 
(${}^3S_1$ : deuteron) and $I=1,J=0$ (${}^1 S_0$) solutions. 
The energy of ${}^3S_1$ is lower than the ${}^1S_0$ because 
$I^{BB}_{11}>I^{AA}_{11}$~(see Table \ref{tab:table3}). 
The order is in agreement with the experimental observations. 
For $B=4$, the quantum number of the ground state $I=0,J=0$ 
coincide with the observation. In experiment, the lowest
excited state also has $I=0,J=0$, which unfortunately can not be explained
 within our scheme. This state can be interpreted as
 ${}^3_2$He+n bound state or ${}^3_1$H+p resonance state 
rather than the resonance of single ${}^4_2$He. Our formulation 
for the multi-soliton is based on the simgle bag-like picture. 
We expect that for some resonance states one needs advanced 
formulation including multi-fragments such as $B=(3+1),(2+2)$. 
We observed $I=0,J=2$ with positive parity as a first 
excited state and this channel should emerge as a higher resonance~(roughly 28 MeV from 
the ground state) in experiment.      
For odd $B$, the constraint of $C_2$ in Eq.~(\ref{j1}) seems 
to assure the validity of the ansatz. Indeed, it provides 
the ground state as $I=J=1/2$ for $B=3$ and as $I=1/2, J=3/2$ for $B=5$, 
which coincide with physical observations. This seems to make 
sense since in the minimal energy configurations with discrete symmetries, 
the solutions tend to have $I=J=1/2$ due to their shell-like 
structure. However, unfortunately the constraint in Eq.~(\ref{j3}) 
forbits such states. 
Consequently, the axially symmetric solitons with odd $B$ emerge 
only as excited states. 
The resultant lowest state is $E=3657$ MeV with $I=1/2,J=5/2$ for $B=3$, 
and $E=6591$ MeV with $I=1/2,J=7/2$ for $B=5$. 
Experimentally, no possible candidate of the state $I=1/2,J=7/2$ are found.  
 
As stated in Sec.~\ref{sec:level2}, in the soliton approach the absolute mass always tends to be higher 
due to the lack of the Casimir effects. Therefore the total energies is overestimated 
around 0.5 $\sim$ 1 GeV in our calculation. 
The one-loop corrections and vacuum effects should be properly subtracted in order to 
estimate physical mass of the solutions. For $B=1$, the obtained mass of the nucleon the 
delta is $E_N=1260$ MeV and $E_\Delta=1505$ MeV respectively because of the lack of 
Casimir effects. With these values, the ground states of the $B=2 \sim 4$ 
exhibit bound states but the ground state of the $B=5$ does not. 
For $B=2$, the state $I=0,J=3$~(${}^7 S_3$) has a large binding 
energy. The search of this resonance is interesting subject and also  
a thorough analysis of the Casimir effects for the toroidal solitons are much 
desired for a stability argument.   

\subsection{$SU(3)$ sector}
The $SU(2)_L \times SU(2)_R$-invariant lagrangian has a natural extension to $SU(3)$ sector 
including strange quarks.  
The $SU(3)_L\times SU(3)_R$-invariant lagrangian in the chiral quark soliton model is given by 
\begin{eqnarray}
\mathcal{L}= 
\bar{\psi}(i\gdhi-MU^{\gamma_5}-\hat{m})\psi
\end{eqnarray}
where $U^{\gamma_5}(x)=e^{i\gamma_5 \pi_a(x)\lambda_a / f_{\pi}}$
and $\lambda_a$ are the usual Gell-Mann matrices with $\lambda_0=\sqrt{\frac{2}{3}}\, \hat{1}$.
In order to estimate the effects of the symmetry breaking of $SU(3)$ explicitly, 
we introduce the current quark mass matrix
\begin{eqnarray}
\hat{m}=\mathrm{diag}(m_0,m_0,m_s)=m_0 \hat{1}+\Delta m (1-\sqrt{3}\lambda_8)/3
\end{eqnarray}
where $\Delta m\equiv m_s-m_0$ is the mass difference between the strange- and the 
u-,d-quark. 

Within the proper-time regularization scheme, we fix our parameters by the input parameters, 
pion mass $m_\pi$, kion mass $m_K$, and the pion decay constant $f_\pi$ in the following 
formulae~\cite{blotz93,wakamatsu96}
\begin{eqnarray}
\frac{N_c M^2}{4\pi^2}\int^\infty_0 \frac{d\tau}{\tau}\phi(\tau)e^{-\tau \bar{M}^2}&=&f^2_{\pi}  
\label{su3_p1}\\
m_0 \frac{N_c M}{2\pi^2 f^2_{\pi}}\int^\infty_0 \frac{d\tau}{\tau^2}\phi(\tau)e^{-\tau \bar{M}^2}&=&m^2_\pi 
\label{su3_p2}\\
(m_0+\frac{\Delta m}{2})\frac{N_c M}{2\pi^2 f^2_\pi}\int^\infty_0  
\frac{d\tau}{\tau^2}\phi(\tau)e^{-\tau \bar{M}^2}&=&m^2_K 
\label{su3_p3}
\end{eqnarray}
where $\bar{M}=M+m_0$. For the damping function $\phi(\tau)$ 
we introduce the two cut-off $\Lambda_1,\Lambda_2$
and the parameter $c$ via 
\begin{eqnarray}
\phi(\tau)=c \theta(\tau-1/\Lambda_1^2)+(1-c) \theta(\tau-1/\Lambda_2^2).
\end{eqnarray}
Using the values $M=400$ MeV , $m_0=6$ MeV , $f_\pi=93$ MeV, $m_\pi=138$ MeV and Eqs.~(\ref{su3_p1}),(\ref{su3_p2})
we obtain the parameter set $\{c=0.76, \Lambda_1=433.8\,\mathrm{MeV}, \Lambda_2=1512.4\,\mathrm{MeV}\}$.
By $m_K=496$ MeV and Eq.~(\ref{su3_p3}), we determine $m_s=149$ MeV. 

For the extension of the three flavor soliton with $B=2$, we follow the usual collective 
coordinate approach for the hedgehog ansatz with $B=1$~\cite{weigel92,blotz93}.
Within the collective coordinate approach, the extension to $SU(3)$ is performed by 
trivial embedding~\cite{kopeliovich90}:
\begin{eqnarray}
\,U(\bm{x},t)=A(t)
\left(
\begin{array}{cc}
U_0(\Lambda^i_j(t) x^j) & 0 \\
0 & 1 \\
\end{array}
\right)
A^{\dagger}(t)\,,
\label{embedding}
\end{eqnarray}
where $A(t)$ is the time-dependent $SU(3)$ collective rotation matrix and 
$\Lambda^i_j(t)=\frac{1}{2}{\mathrm Tr} (\tau^i B \tau_j B^\dagger)$ is the 
spatial rotation matrix. Substituting (\ref{embedding}) into the quark 
determinant and transforming the rotated frame of reference, 
one obtains~\cite{sawado02}: 
 \begin{eqnarray}
&&i D= i \gdhi-M 
    U^{\gamma_5}(\bm{x},t)-\hat{m} \nonumber \\
&&\rightarrow A(t) S(t)^{\dagger}\gamma_0(i\partial_t-H(U^{\gamma_5}_S) 
-H_{SB}-\Omega_A+\Omega_B)S(t)A(t)^{\dagger}\,,
\label{su3_doperator}
\end{eqnarray}
where
\begin{eqnarray} 
&&\Omega_A =-i A^{\dagger}\dot{A}=\frac{1}{2}\Omega_A^a \lambda_a\,, \\  
&&\Omega_B=\Omega_B^a (\frac{1}{2}\epsilon_{abc}\gamma^b \gamma^c -i(\bm{r}\times\nabla)_a)
=\Omega_B^a J_a\,,
\end{eqnarray}
and 
\begin{eqnarray}
H_{SB}=A^{\dagger}(t)\beta\Delta m \frac{1}{3}(1-\sqrt{3}\lambda_8) A(t)\,.
\end{eqnarray}
$S(t)$ is the rotation operator for the Dirac fields. In the rotating system, 
the quarks feel the induced Coliolis forces $\Omega_A, \Omega_B$ and $H_{SB}$. 
$\Omega_A, \Omega_B$ are the angular velocity operators for the right flavor 
rotation and the spatial rotation.
The $H_{SB}$ represents the contribution to the hamiltonian due to the $SU(3)$ 
symmetry breaking. 

We assume that the rotational velocities and the mass difference are relatively small 
and the expansion in powers of $\Omega_A, \Omega_B$ and $\Delta m$ is rapidly convergent. 
Expanding the quark determinant and the (valence quark) Green function in terms of the 
collective angular velocity up to the second order and the quark mass 
difference of the first order. The effective action~(\ref{effective_det}) 
with Eq.~(\ref{su3_doperator}) can be
rewrite as
\begin{eqnarray}
S_{\rm eff}=-iN_c {\rm Sp} \log iD \to S_{\rm eff}(U_S)+S_R+S_I\,.
\end{eqnarray}
$S_R,S_I$ mean the real and imaginary contribution to the action, which are explicitly written as
\begin{eqnarray}
S_R=-\frac{i}{2}N_c[{\rm Sp}\log dd^\dagger-{\rm Sp}\log d_0{d_0}^\dagger]\,,~~
S_I=-\frac{i}{2}N_c{\rm Sp}\log d(d^\dagger)^{-1}\,,
\end{eqnarray}
where $d=i\partial_t-H(U_S)-H_{\rm SB}-\Omega_A+\Omega_B$.
The expansion for the real part with proper-time regularization form yields 
\begin{eqnarray}
S_R&\to& -\int \frac{d\tau}{\tau}{\rm Sp}\phi(\tau)[e^{-dd^\dagger\tau}-e^{-d_0{d_0}^\dagger\tau}] \nonumber \\
&\to& \int dt \biggl[-\frac{1}{2}\gamma_0 (1-D_{88})
+\frac{1}{2}I^{AA}_{ab,0}\Omega^a_A \Omega^b_A
-I^{AB}_{ab,0}\Omega^a_A \Omega^b_B
+\frac{1}{2}I^{BB}_{ab,0}\Omega^a_B \Omega^b_B \biggr]   \nonumber \\
\end{eqnarray}
and 
\begin{eqnarray}
&&\gamma_0=-\Delta m\frac{N_c}{3}\sum_{i=1,2} \sum_{n}c_i~\mathrm{sgn}(E_n)\mathrm{erfc}
\Bigl(\frac{|E_n|}{\Lambda_i}\Bigr) \langle n|\gamma_0 |n\rangle\,,   \\
&&I^{AA}_{ab,0}=\frac{N_c}{8}\sum_{i=1,2}\sum_{m,n}c_i~f(E_m,E_n,\Lambda_i) \langle n | \lambda_a | m\rangle 
                     \langle m | \lambda_b | n \rangle \,,   \\
&&I^{AB}_{ab,0}=\frac{N_c}{4}\sum_{i=1,2}\sum_{m,n}c_i~f(E_m,E_n,\Lambda_i) \langle n | \lambda_a | m\rangle 
                     \langle m | J_b | n \rangle \,,   \\
&&I^{BB}_{ab,0}=\frac{N_c}{2}\sum_{i=1,2}\sum_{m,n}c_i~f(E_m,E_n,\Lambda_i) \langle n | J_a | m\rangle 
                     \langle m | J_b | n \rangle\,,  
\end{eqnarray}
where $c_1=c, c_2=1-c$ and the cutoff function
 $f(E_m,E_n,\Lambda)$ is defined in Eq.~(\ref{cutoff}).
For the estimation of the imaginary part of the action, 
we need the following manipulation, 
\begin{eqnarray}
S_I&=&-\frac{i}{2}N_c{\rm Sp}\log d(d^\dagger)^{-1} \nonumber \\
&=&\int dt \int \frac{d\omega}{2\pi}
{\rm Tr} \log \biggl(\frac{w-H(U_S)-H_{SB}-\Omega_A+\Omega_B}
{w-H(U_S)-H_{SB}+\Omega_A-\Omega_B}\biggr)
 \nonumber \\
&=&\int dt \int \frac{d\omega}{2\pi} \int^1_{-1} d\lambda
{\rm Tr} \biggl(\frac{-\Omega_A+\Omega_B}
{w-H(U_S)-H_{SB}-\lambda(\Omega_A-\Omega_B)}\biggr)\,. \nonumber 
\end{eqnarray}
With replacement $\omega-\lambda(\Omega_A-\Omega_B) \to \omega$ and analytical 
continuation for $\omega$, we obtain
\begin{eqnarray}
&=&\int dt \int \frac{d\omega}{2\pi} \int^1_{-1} d\lambda
{\rm Tr} \biggl(\frac{-\Omega_A+\Omega_B}
{w-H(U_S)-H_{SB}}\biggr) \nonumber \\
&=&2i \int \frac{d\omega'}{2\pi}{\rm Sp}
\frac{(\Omega_A-\Omega_B)(H(U_S)+H_{SB})}{\omega'^2+(H(U_S)+H_{SB})^2}\,,~~\omega=i\omega'\,.
\end{eqnarray}
Introducing the following proper-time regularization,
\begin{eqnarray}
\frac{1}{\omega'^2+(H(U_S)+H_{SB})^2}\to \int^\infty_{1/\Lambda^2}d\tau
\exp[-\tau(\omega'^2+(H(U_S)+H_{SB})^2)]
\end{eqnarray}
and expand up to first order, we finally obtain the form
\begin{eqnarray}
S_I\to-\int dt\biggl[\frac{\sqrt{3}}{2} B [U_S] \Omega^8_A
+K^A_{ab,0}D_{8a}\Omega_A^b+K^B_{ab,0}D_{8a}\Omega_B^b \biggr]
\end{eqnarray}
and
\begin{eqnarray}
&&K^A_{ab,0}=\Delta m\frac{N_c}{4\sqrt{3}} \sum_{m,n}F(E_m,E_n,\Lambda) 
 \langle n | \beta \lambda_a | m\rangle 
\langle m | \lambda_b | n \rangle \,,   \\
&&K^B_{ab,0}=\Delta m\frac{N_c}{2\sqrt{3}} \sum_{m,n}F(E_m,E_n,\Lambda)
 \langle n | \beta \lambda_a | m\rangle 
\langle m | J_b | n \rangle \,.   
\end{eqnarray}
The moments of inertia $K^A_{ab}, K^B_{ab}$ are derived from imaginary part of the effective
action thus need no regularization. The ``cut-off'' function $F(E_m,E_n,\Lambda)$ 
becomes
\begin{eqnarray}
F(E_m,E_n,\Lambda\to\infty) 
=\frac{{\rm sgn}(E_m)-{\rm sgn}(E_n)}{E_m-E_n}\,. 
\end{eqnarray}
Also the valence quark contributions for the moments of inertia read
\begin{eqnarray}
&&\gamma_{\rm val}=\Delta m\frac{2 N_c}{3} 
\langle {\rm val}|\gamma_0 |{\rm val} \rangle\,,  \\
&&K^A_{ab,{\rm val}}=\Delta m\frac{N_c}{\sqrt{3}} \sum_{n\ne {\rm val}}
\frac{\langle {\rm val} | \beta \lambda_a | n\rangle 
\langle n | \lambda_b | {\rm val} \rangle}{E_n-E_{\rm val}} \,,   \\
&&K^B_{ab,{\rm val}}=\Delta m\frac{2N_c}{\sqrt{3}} \sum_{n\ne {\rm val}}
\frac{\langle {\rm val} | \beta \lambda_a | n\rangle 
\langle n | J_b | {\rm val} \rangle}{E_{n}-E_{\rm val}} \,,  \\
&&I^{AA}_{ab,{\rm val}}=\frac{N_c}{2}\sum_{n\ne {\rm val}}
\frac{\langle {\rm val}| \lambda_a | n\rangle 
\langle n | \lambda_b | {\rm val} \rangle}{E_n-E_{\rm val}} \,,  \\
&&I^{AB}_{ab,{\rm val}}=N_c\sum_{n\ne {\rm val}}\frac{\langle {\rm val} | \lambda_a | n\rangle 
\langle n | J_b | {\rm val} \rangle}{E_n-E_{\rm val}} \,,  \\
&&I^{BB}_{ab,{\rm val}}=2 N_c\sum_{n\ne {\rm val}}\frac{\langle {\rm val} | J_a | n\rangle 
\langle n | J_b | {\rm val} \rangle}{E_n-E_{\rm val}} \,. 
\end{eqnarray}

Finally we can construct the following form for the effective lagrangian 
\begin{eqnarray}
L&=&-E_{\rm static}-\frac{\sqrt{3}}{2} B [U_S] \Omega^8_A-\frac{1}{2}\gamma (1-D_{88}) \nonumber  \\ 
&-&K^A_{ab}D_{8a}\Omega_A^b-K^B_{ab}D_{8a}\Omega_B^b \nonumber \\
&+&\frac{1}{2}I^{AA}_{ab}\Omega^a_A \Omega^b_A
-I^{AB}_{ab}\Omega^a_A \Omega^b_B
+\frac{1}{2}I^{BB}_{ab}\Omega^a_B \Omega^b_B\,,
\label{su3_lagrangian}
\end{eqnarray}
and total moments of inertia acquire due to their sum, e.g.
 $I^{AA}_{ab}=I^{AA}_{ab,{\rm val}}+I^{AA}_{ab,0}$.
where $E_{\rm static}$ is the self-consistent classical soliton energy and
 $D_{ab}(A)=\frac{1}{2}{\rm Tr}
 (\lambda_a A \lambda_b A^\dagger)$ is a $SU(3)$ Wigner rotation matrix. 

In the evaluations of the moments of inertia, the numerical difficulties may arise. Due to the 
difference of the boundary conditions between the initial and the final states of the matrix 
elements one may obtain the spurious nonzero values of the moments of inertia in the absence of 
background pion fields. Similar problem occurs in the $\langle n|J_a| m \rangle$ for 
all values of $a$ and $\langle n|\lambda_a| m \rangle$ for $a \ge 4$ with our harmonic 
oscillator basis. To avoid the problem, we employ the 
replacement of Eq.~(\ref{eq:replacement}) for $\langle n|J_a| m \rangle$ \cite{sawado02}. 
 For the matrix element of $\lambda_a$, similar replacement induces an additional spurious term 
proportional to the mass difference.  However, within our perturbative treatment in the mass 
difference, this procedure is justified. 

With the standard canonical quantization formula for the collective coordinates we obtain the 
following quantization prescriptions
\begin{eqnarray}
&&R_a=
\left\{\begin{array}{l}
-\sum_j(I^{AA}_{aj}\Omega^j_A-I^{AB}_{aj}\Omega^j_B-K^A_{aj}D_{8j})\,, 
\hspace{0.3cm}a=1,2,3, \\
-\sum_b(I^{AA}_{a b}\Omega^b_A-K^A_{ab}D_{8b})\,,  
\hspace{2cm}a=4,5,6,7, \\
\frac{\sqrt{3}}{2} B\,,  
\hspace{5.3cm} a=8,
\end{array}\right.
\end{eqnarray}
and
\begin{eqnarray}
K_i=-\sum_j(I^{BB}_{ij}\Omega^j_B-I^{BA}_{ij}\Omega^j_A+K^B_{ij}D_{8j})\,,~~i=1,2,3,
\end{eqnarray}
where $R_a$ is the right isospin generator of $SU(3)$, and $K_i$ represents the generator 
of the spatial rotation. 

Due to the symmetry of the soliton, only the following elements of the moments of inertia
survive: 
\begin{eqnarray}
&&I_{11}^{AA}=I_{22}^{AA},I_{11}^{BB}=I_{22}^{BB},\nonumber \\ 
&&I_{33}^{BB}=m^2I_{33}^{AA}, I_{33}^{AB}=I_{33}^{BA}=-mI_{33}^{AA}, \nonumber \\
&&I_{44}^{AA}=I_{55}^{AA}=I_{66}^{AA}=I_{77}^{AA}, \nonumber \\
&&K_{11}^{AA}=K_{22}^{AA},\nonumber \\ 
&&K_{33}^{BB}=-mK_{33}^{AA}, \nonumber \\
&&K_{44}^{AA}=K_{55}^{AA}=K_{66}^{AA}=K_{77}^{AA}\,.
\end{eqnarray}
Thus the hamiltonian becomes
\begin{eqnarray}
H&=&E_{\rm static}+H_0+H_1, \, \nonumber \\
H_0&=&\frac{1}{2}\frac{1}{I^{AA}_{44}}\sum^{7}_{a=1}R_a^2
+\frac{1}{2}(\frac{1}{I^{AA}_{11}}-\frac{1}{I^{AA}_{44}})\sum^{3}_{i=1}R_i^2 \nonumber \\
&+&\frac{1}{2}\frac{1}{I^{BB}_{11}}\sum^{3}_{i=1}K_i^2
+\frac{1}{2}(\frac{1}{I^{AA}_{33}}-\frac{1}{I^{AA}_{11}}-\frac{m^2}{I^{BB}_{11}})R^2_3\,.
\label{su3_hamiltonian}
\end{eqnarray}
In the evaluation of the $H$, we adopt a simple perturbative treatment with 
the mass difference $\Delta m$~\cite{blotz93}. 
Up to first order of the $\Delta m$, the $H_1$ is written as
\begin{eqnarray}
H_1&=&\frac{1}{2}\gamma(1-D_{88})
-\frac{K^A_{11}}{I^{AA}_{11}}(D_{81}R_1+D_{82}R_2)  \nonumber \\
&-&\frac{K^A_{33}}{I^{AA}_{33}}D_{83}R_3 
-\frac{K^A_{44}}{I^{AA}_{44}}\sum^7_{k=4}D_{8k}R_k\,.
\end{eqnarray}
The hamiltonian in Eq.~(\ref{su3_hamiltonian}) is diagonalized by using  
following collective wave functions of the nonperturbative part of the hamiltonian $H_0$
\begin{eqnarray}
&&\Psi^{(n)}_{YII_3,Y'NN_3,JJ_3}(A,B) 
=\sqrt{\mathrm{dim}(n)}(-1)^{\frac{Y'}{2}+N_3} \nonumber \\
&&\times{D^{(n)}}^{*}_{YII_3,Y'NN_3}(A){D^{J}}^{*}_{J_3,-mN_3}(B)\,.
\label{su3_wf}
\end{eqnarray}
With these bases, the matrix element reduces to the integral of the three
Wigner matrices which can be easily evaluated from the $SU(3)$ Clebsch-Gordan 
coefficients~\cite{blotz93,toyota87}. 
The actual computations of the Clebsch-Gordan coefficients can be performed by using 
the numerical algorithm in Ref.~\cite{kaeding96}. 

\begin{table}[htb]
\begin{center}
\caption{\label{tab:su3_inrtia}The various moments of inertia, with $m_s=149$ MeV.}
\newcommand{\m}{\hphantom{$-$}}
\newcommand{\cc}[1]{\multicolumn{1}{c}{#1}}
\renewcommand{\tabcolsep}{0.4pc} 
\renewcommand{\arraystretch}{1.1} 
\begin{tabular}{@{}cccc} \hline
         & Valence & Vacuum & Total  \\  \hline
$I^{AA}_{11} [\mathrm{GeV}^{-1}]$ & $7.49 $ & $4.05 $ & $11.54 $ \\
$I^{BB}_{11} [\mathrm{GeV}^{-1}]$ & $11.19 $ & $5.46 $ & $16.65 $ \\
$I^{AA}_{33} [\mathrm{GeV}^{-1}]$ & $4.36 $ & $1.61 $ & $5.97 $ \\
$I^{AA}_{44} [\mathrm{GeV}^{-1}]$ & $1.64 $ & $1.26 $ & $2.90 $ \\
$K^{A}_{11}$ & $0.285 $ & $1.38\times10^{-4}$ & $0.285 $ \\
$K^{A}_{33}$ & $0.297 $ & $1.60\times10^{-3}$ & $0.298 $ \\
$K^{A}_{44}$ & $0.255 $ & $-1.05\times10^{-3}$ & $0.254 $ \\
$\gamma [\mathrm{MeV}]$ &292.0 & 1098.8 & 1390.8 \\ \hline
\end{tabular}
\end{center}
\end{table}

Our numerical calculations were performed with the constituent mass $M=400$ MeV. 
For a diagonalization problem of the Dirac hamiltonian, 
we used the deformed harmonic oscillator basis~\cite{gambhir} which is described in detail in Appendix A. 
The self-consistent classical mass was obtained as $E_{\rm static}=2406$ MeV which differs 
from the result of $SU(2)$ sector. The difference arises due to the specific choice of 
the cutoff scheme in Eqs.~(\ref{su3_p1})-(\ref{su3_p3}).  
The values of various moments of inertia are listed in Table \ref{tab:su3_inrtia}. 
The quantized states are coupled to the 
multiplets $\{\overline{10} \}$, $\{27 \}$, $\{35 \}$, $\{28 \}$ corresponding to 
$(p,q)=(0,3),(2,2),(4,1),(6,0)$ respectively. In Table \ref{tab:su3_spectra}, 
we show all the mass of the dibaryon states for the multiplets. 
The energy levels of the dibaryon states belonging to each multiplets are 
shown in Fig.~\ref{fig:su3_spectra}. 
\begin{table}[htb]
\begin{center}
\caption{\label{tab:su3_spectra}Absolute mass of the dibaryon (in MeV), with $m_s=149$ MeV.}
\newcommand{\m}{\hphantom{$-$}}
\newcommand{\cc}[1]{\multicolumn{1}{c}{#1}}
\renewcommand{\tabcolsep}{0.75pc} 
\renewcommand{\arraystretch}{1.1} 
\begin{tabular}{@{}ccccccc}\hline                         
 Multiplet &($S$ $I$ $J$) & Mass & Multiplet & ($S$ $I$ $J$) & Mass\\  \hline
$\overline{10}$ & (\,0\,  0\,  1)  & 3255 & $35$ &(\,0\,  2\,  1)         & 3610 \\
                &                  &      &      &(-1  $\frac{5}{2}$  1)  & 3965 \\
            &(-1  $\frac{1}{2}$ 1) & 3467 &      &(-1  $\frac{3}{2}$  1)  & 3727 \\
            &                      &      &      &(-2  2  1)              & 4034 \\
            &(-2  1  1)            & 3679 &      &(-2  1  1)              & 3844 \\
            &                      &      &      &(-3  $\frac{3}{2}$  1)  & 4103 \\
            &(-3  $\frac{3}{2}$  1)& 3891 &      &(-3  $\frac{1}{2}$  1)  & 3960 \\
            &                      &      &      &(-4  1  1)              & 4172  \\ 
            &                      &      &      &(-4  0  1)              & 4077  \\
$27$ & (\,0\,   1\,    0)     & 3309 &           &(-5  $\frac{1}{2}$  1)  & 4241  \\
     & (-1  $\frac{3}{2}$  0) & 3573 &           &                        &       \\
     & (-1  $\frac{1}{2}$  0) & 3453 &      $28$ &(\,0\,  3\,  0)         & 3969  \\
     & (-2  2  0)             & 3841 &           &(-1  $\frac{5}{2}$  0)  & 4054  \\ 
     & (-2  1  0)             & 3678 &           &(-2  2  0)              & 4139  \\
     & (-2  0  0)             & 3597 &           &(-3  $\frac{3}{2}$  0)  & 4224  \\
     & (-3  $\frac{3}{2}$  0) & 3904 &           &(-4  1  0)              & 4309  \\
     & (-3  $\frac{1}{2}$  0) & 3781 &           &(-5  $\frac{1}{2}$  0)  & 4393  \\ 
     & (-4  1  0)             & 3966 &           &(-6  0  0)              & 4478  \\
 \hline
\end{tabular}
\end{center}
\end{table}

In a pioneering work of the $SU(3)$ collective quantization of the chiral soliton~\cite{kunz88}, 
it is pointed out that because of the constraint $Y_R=2$ which arises from the trivial embedding, 
some states in the constituent quark model are not allowed in the soliton solution. 
Hence the state we obtained is not the lowest state in the $S=-2$ sector and 
the configuration of H-dibaryon may have not be an axially symmetric.  
On the other hand, $(I,J)=(0,0)$ channel in the $S=-6$ sector, corresponding to di-Omega $\Omega \Omega$ 
could have a rather deeper bound. According to the data of the $B=1$ hedgehog analysis
 (in Ref.~\cite{blotz93}), 
we expect the binding energy about $\sim 200$ MeV.  This state is rather promising 
as a candidate of the axially symmetric dibaryon.

In the perturbative treatment of Eqs.~(\ref{su3_lagrangian}),(\ref{su3_hamiltonian}) 
we retained only linear terms for the mass difference $\Delta m$ and  
used $SU(3)$ symmetric wave functions in Eq.~(\ref{su3_wf}) for the ground state.
For the $B=1$ hedgehog case, such leading order approximation induces discrepancy 
with experiment~\cite{blotz93}. But it is possible to minimize it by incorporating 
appropriate higher-order effects in the $\Delta m$~\cite{blotz96}.
In our estimation of moments of inertia, in order to remove the spurious contributions, 
we employ ad hoc approximation for the matrix elements and it is justified 
only if we confine our calculations up to first order. 
We expect an extension of our scheme to the second order in $\Delta m$ is also feasible.
 
\section{\label{sec:level6}Concluding Remarks}
This article reports the theoretical framework and the numerical results of the 
chiral quark soliton model for higher baryon number solutions. 
The model was inspired by the instanton liquid model of the QCD vacuum 
and thus incorporates the basic features of QCD, e.g. the chiral symmetry 
and its breakdown accompanied by the appearance of the Goldstone bosons. 
This model provides correct observable as a nucleon including mass, 
electromagnetic value, spin carried by quarks, parton distributions
and octet, decuplet $SU(3)$ baryon spectra.
For $B>1$, different topological configurations from $B=1$ 
are needed to obtain the minimal energy solutions. For $B=2$, 
we employed the axially symmetric configuration which produces the minimal 
energy configuration in the Skyrme model. 
We also investigated the $B=3,4,5$ solitons with axial symmetry. 
The solution exhibits doubly degenerate bound spectra of the 
one-quark orbits. This relatively large degeneracy confirms 
that the solutions are stable local minima. For $B>2$, the 
Skyrme model predicts that the solutions have only discrete
symmetries. According to the prediction, we studied the CQSM
with the chiral fields of such platonic symmetries. The discrete 
crystal-like symmetries exhibit much complicated structure and 
the study of such configurations is rather formidable task. 
However the analysis becomes much simpler when we adopt the rational 
map ansatz to the chiral fields since with this ansatz the chiral fields 
are separable in polar coordinates and radial coordinate, which 
makes the numerical technique developed for $B=1$ applicable to find solutions with 
higher $B$.
We showed that the baryon densities inherit the same discrete symmetries 
as the chiral fields and obtained various degenerate bound spectra of the valence 
quarks depending on the background of chiral field configurations. 
Evaluating the radial component of the baryon density, shell-like structure of the 
valence quark spectra was observed. 
The group theory should predict these level structures resulting from the symmetry
of the background potential. In fact the degeneracy of the valence spectra 
are determined by the winding number of the chiral fields as well as the shape 
deformation (symmetry) of solitons. The four-fold degeneracy of the lowest states may be 
ascribed to the chiral symmetry $SU(2)_L\times SU(2)_R$ of the hamiltonian. 
To get better understanding of the relation between the quark level structure and 
the winding number or the shape deformation, further analysis will be worth to be done 
in future.  

Upon quantization, we computed zero-mode rotational corrections to 
the classical energy. The study of the Finkelstein-Rubinstein 
constraints indicates that the axially symmetric solution with 
even $B$ has the same quantum number as the physically observed 
nuclei. Some of the states may be observed in experiments. 
For example, in the $B=2$, we obtained $I=0,J=1$ 
(${}^3S_1$ : deuteron) and $I=1,J=0$ (${}^1 S_0$) solutions. 
The energy of ${}^3S_1$ is lower than the ${}^1S_0$. 
The order is in agreement with the experimental observations. 
For $B=4$, the quantum number of the ground state $I=0,J=0$ 
coincide with the observation. For odd $B$, 
the constraint of $C_2$ in Eq.~(\ref{j1}) seems 
to assure the validity of the ansatz. Indeed, it provides 
the ground state as $I=J=1/2$ for $B=3$ and as $I=1/2, J=3/2$ for $B=5$ 
due to their shell-like structure, which coincide with physical observations. 
However, unfortunately the constraint in Eq.~(\ref{j3}) forbid such states. 
Consequently, the axially symmetric solitons with odd $B$ can appear
only as excited states. 
   
For an $SU(3)$ extension of the model, we adopted the collective quantization 
scheme with a trivial embedding form for the chiral fields. In order to 
estimate the effects of the quark mass difference, we performed the naive perturbative method 
in terms of the mass difference. We obtained the dibaryonic spectrum coupled to the multiplets
of $\{\overline{10} \}$, $\{27 \}$, $\{35 \}$, $\{28 \}$ . 
The state we obtained is not the lowest state in the $S=-2$ sector. 
The configuration of H-dibaryon may have not be an axially symmetric.  
On the other hand, $(I,J)=(0,0)$ channel in the $S=-6$ sector, corresponding to di-Omega $\Omega \Omega$ 
may have a rather deeper bound. This state is promising as a candidate of the axially symmetric dibaryon.

In our analysis, all obtained states seem to be deep bound states. 
Consideration of the Casimir effect is, however, necessary to determine which
states are stable.  
For the $B=1$ skyrmion, the Casimir energies of the rotational and the translational 
zero modes were estimated by various authors. Their predictions for the total mass are around 
 $-0.5 \sim -1.3$ GeV~\cite{moussallam91,moussallam93,holzwarth,weigel95,scholtz}. 
For the $B=2$ torus, the Casimir energies have not been estimated yet. 
The thorough analysis of the Casimir effects is desired in order to examine the stability. 

\appendix
\section{Numerical Basis}
\subsection{Kahana-Ripka basis}
The numerical method widely used in this model is 
based on the expansion of the Dirac spinor in terms of an appropriate 
orthogonal basis. 
The Kahana-Ripka basis~\cite{kahana84} which was originally constructed for 
diagonalizing the 
hamiltonian with the chiral fields of $B=1$ hedgehog ansatz is a plane-wave 
finite basis. The basis is discretized by imposing an appropriate boundary 
condition on the radial wave functions at the radius $r_{\rm max}$ chosen 
to be sufficiently larger than the soliton size. 
The basis is then made finite by including only those states 
with the momentum $k$ as $k<k_{\rm max}$. 
The results should be, however, independent on $r_{\rm max}$ and $k_{\rm max}$.

The hamiltonian with hedgehog ansatz commutes with the parity and the grandspin operator 
given by  
\begin{eqnarray*}
	\bm{K}=\bm{j}+\bm{\tau}/2=\bm{l}+\bm{\sigma}/2+\bm{\tau}/2,
\end{eqnarray*}
where $\bm{j},\bm{l}$ are respectively total angular momentum and orbital angular momentum. 
Accordingly, the angular basis can be written as  
\begin{eqnarray}
|(lj)KM\rangle= \sum_{j_3\tau_3}C^{KM}_{jj_3\frac{1}{2}\tau_3}
\Bigl(\sum_{m\sigma_3}C^{jj_3}_{lm\frac{1}{2}\sigma_3}
|lm \rangle |\frac{1}{2}\sigma_3 \rangle \Bigr) |\frac{1}{2} \tau_3 \rangle\,.
\label{angularbasis}
\end{eqnarray}
With this angular basis, the normalized eigenstates of the free hamiltonian 
in a spherical box with radius $r_{\rm max}$ can be constructed as follows:
\begin{eqnarray}
&&u^{(1)}_{KM}=
N_k\left( 
\begin{array}{c}
ij_{K}(kr)|(K K+\frac{1}{2})KM\rangle \\
\Delta_k j_{K+1}(kr)|(K+1 K+\frac{1}{2})KM\rangle
\end{array}
\right), \nonumber \\
&&u^{(2)}_{KM}=
N_k\left( 
\begin{array}{c}
ij_{K}(kr)|(K K-\frac{1}{2})KM\rangle \\
-\Delta_k j_{K-1}(kr)|(K-1 K-\frac{1}{2})KM\rangle
\end{array}
\right), \nonumber \\
&&u^{(3)}_{KM}=
N_k\left( 
\begin{array}{c}
i\Delta_k j_{K}(kr)|(K K+\frac{1}{2})KM\rangle \\
-j_{K+1}(kr)|(K+1 K+\frac{1}{2})KM\rangle
\end{array}
\right), \nonumber \\
&&u^{(4)}_{KM}=
N_k\left( 
\begin{array}{c}
i\Delta_kj_{K}(kr)|(K K-\frac{1}{2})KM\rangle \\
j_{K-1}(kr)|(K-1 K-\frac{1}{2})KM\rangle
\end{array}
\right), \nonumber \\
\nonumber \\
&&v^{(1)}_{KM}=
N_k\left( 
\begin{array}{c}
ij_{K+1}(kr)|(K+1 K+\frac{1}{2})KM\rangle \\
-\Delta_k j_{K}(kr)|(K K+\frac{1}{2})KM\rangle
\end{array}
\right), \nonumber \\
&&v^{(2)}_{KM}=
N_k\left( 
\begin{array}{c}
ij_{K-1}(kr)|(K-1 K-\frac{1}{2})KM\rangle \\
\Delta_k j_{K}(kr)|(K K-\frac{1}{2})KM\rangle
\end{array}
\right), \nonumber \\
&&v^{(3)}_{KM}=
N_k\left( 
\begin{array}{c}
i\Delta_k j_{K+1}(kr)|(K+1 K+\frac{1}{2})KM\rangle \\
j_{K}(kr)|(K K+\frac{1}{2})KM\rangle
\end{array}
\right), \nonumber \\
&&v^{(4)}_{KM}=
N_k\left( 
\begin{array}{c}
i\Delta_kj_{K-1}(kr)|(K-1 K-\frac{1}{2})KM\rangle \\
-j_{K}(kr)|(K K-\frac{1}{2})KM\rangle
\end{array}
\right), \label{kahana_ripka}
\end{eqnarray}
with
\begin{eqnarray}
	N_k=\biggl[\frac{1}{2}r_{\rm max}^3
	\Bigl(j_{K+1}(kr_{\rm max})\Bigr)^2\biggr]^{-1/2}
\end{eqnarray}
and $\Delta_k=k/(E_k+M)$. 

The momenta are discretized by the 
boundary condition $j_K(k_i r_{\rm max})=0$.
The  $u,v$ correspond to the {\it ``natural''} 
and {\it ``unnatural''} components of the basis  
which stand for parity $(-1)^{K}$ and $(-1)^{K+1}$ respectively. 

Let us construct the trial function using the Kahana-Ripka basis 
to solve the eigenequations in Eq.~(\ref{eigen_h}), 
\begin{eqnarray}
	&&\phi_\mu(\bm{x})=\lim_{K_{\rm max}\to\infty}\sum^{K_{\rm max}}_{K=0}
      \sum^{K}_{M=-K}\sum_{j=1}^{4}[\alpha_{KM,\mu}^{(j)}u_{KM}^{(j)}(r,\theta,\varphi)\nonumber \\
      &&\hspace{3cm}+\beta_{KM,\mu}^{(j)}v_{KM}^{(j)}(r,\theta,\varphi)]	
	\label{expand_basis}.
\end{eqnarray}

\subsection{Deformed Oscillator basis}
Let us show the numerical analysis of the eigen equations in detail.  
To solve the eigenequation of the form,  
\begin{eqnarray}
	&&[-i \bm{\alpha}\cdot\nabla + \beta M (\cos F(\rho,z)  + i {\gamma}_{5} 
	\bm{\tau} \cdot \hat{\bm{n}}_R \sin F(\rho,z)]\phi_\mu(\bm{x})=E_\mu\phi_\mu(\bm{x})\,,
	\label{eigenequation2}
\end{eqnarray}
we introduce the deformed harmonic oscillator spinor basis which was 
originally constructed by Gambhir {\it et al.} in the relativistic 
mean field theory for deformed nuclei~\cite{gambhir}. 
The upper and lower components of the Dirac spinors are 
expanded separately by the basis as 
\begin{equation}
	{\phi}_{\mu} (\bm{x})= 
	\left( 
	\begin{array}{c}
	f_{\mu} (\bm{x}) \\
	ig_{\mu} (\bm{x}) 
	\end{array}\right)\; 
	 =  \left( 
	\begin{array}{c}
	\sum_{a}f_{\mu a} \bm{\Phi}_{a} (\bm{x},s) \\
	i \sum_{ \tilde{a}}g_{\mu \tilde{a}} \bm{\Phi}_{\tilde{a}} (\bm{x},s)  
	\end{array}\right)\; 
	{\chi}^I_{m_{\tau}}
\end{equation}
where $\bm{\Phi}_{a} (\bm{x},s,\tau) $ are the eigefunctions of 
a deformed harmonic oscillator potential
\begin{eqnarray}
	V_{osc}(\rho,z)=\frac{1}{2}{\mathcal M}\omega^2_\rho \rho^2
	+\frac{1}{2}{\mathcal M}\omega^2_z z^2 \,,
\end{eqnarray}
and defined by   
\begin{equation}
	\bm{\Phi}_{a} (\bm{x},m_s) = \frac{1}{\sqrt{2 \pi}} \phi^{|\omega|}_{n_{r}}({\rho}) 
	{\phi}_{n_{z}}(z) e^{i\omega \varphi} {\chi}^S_{m_{s}}
	\label{deformed_basis}
\end{equation}
with
\begin{eqnarray*}
	&&{\phi}^{|\omega|}_{n_{r}}({\rho}) = N_{n_r}^{|\omega|}
	(\sqrt{{\alpha}_{\rho} \rho})^{|\omega|} e^{- \frac{1}{2} 
	{\alpha}_{\rho} {\rho}^{2}}L^{|\omega|}_{n_r}(\alpha_\rho \rho^2) \\
	&&\hspace{1cm}n_{r}=0,1,2, \cdot \cdot \cdot, N_{\rm rmax} \\
	&&{\phi}_{n_z}(z) = N_{n_z}
	 e^{- \frac{1}{2} \alpha_z z^2}H_{n_z}
	(\sqrt{{\alpha}_{z}} {z}) \\
	&&\hspace{1cm}n_{z}=1,3, \cdot \cdot \cdot, 2N_{\rm zmax}+1~~{\rm or}~~0,2,\cdots 
	, 2N_{\rm zmax} \,,
\end{eqnarray*}
and 
\begin{equation}
	{\chi}_{+}=\left( 
	\begin{array}{c}
	1  \\
	0
	\end{array}\right)~,~~~~~~{\chi}_{-}=\left( 
	\begin{array}{c}
	0  \\
	1
	\end{array}\right) 
\end{equation} 
depending on if the eigenvalues of the third components 
of the spin $m_{s}$ (isospin $m_{\tau}$) takes $+1$ or $-1$. 
The functions, $L^{|m|}_{n_{r}}$ and 
$H_{n_{z}}$, are the associated Laguerre polynomials 
and the Hermite polynomials with the normalization constants
\begin{eqnarray}
	N_{n_r}^{|\omega|}=\sqrt{\frac{2\alpha_\rho n_r!}{(n_r+|\omega|)!}}\,\,,\,\,
	N_{n_z}=\frac{1}{\sqrt{2^{n_z}n_z!\sqrt{\frac{\pi}{\alpha_z}}}}\,.
\end{eqnarray}
These polynomials can be calculated by following recursion relations
\begin{eqnarray}
	x\frac{d}{dx}L^{m}_{n}(x)=nL^{m}_{n}(x)-(n+m)L^{m}_
	{n-1}(x) \\
	L^{m-1}_{n}(x)=L^{m}_{n}(x)-L^{m}_{n-1}(x)
\end{eqnarray}
and
\begin{eqnarray}
	H_{n+1}(x)-2xH_{n}(x)+2n_zH_{n-1}(x)=0\\
	\frac{d}{dx}H_{n}(x)=2nH_{n-1}(x) 
\end{eqnarray}
where constants ${\alpha}_{\rho}$ and ${\alpha}_{z}$ can be expressed
by the oscillator frequencies as
\begin{eqnarray}
\alpha_\rho=\frac{{\mathcal M}\omega_\rho}{\hbar}\,,\,\, 
\alpha_z=\frac{{\mathcal M}\omega_z}{\hbar}
\end{eqnarray}
which are free parameters chosen optimally. 
The $N_{\rm rmax}$ and $N_{\rm zmax}$ are increased until convergence is attained. 
The parity transformation rule of ${\Phi}_{\alpha}$ is given by 
\begin{equation}
	{\Phi}_{\alpha}(\rho,\varphi + \pi ,-z;s,t)={(-1)}^{\omega+{n}_{z}}{\Phi}_{\alpha}
	(\rho,\varphi,z;s,t) 
\end{equation}
where 
\begin{equation}
	H_{n_z} (-\sqrt{\alpha_z} z)={(-1)}^{n_z}{H}_{n_z} 
	(\sqrt{\alpha_z} z),
	\label{eq:61}
\end{equation}
has been used.
The parity is $+$ for $\omega+{n}_{z}=$ odd, and  
$-$ for $\omega+{n}_{z}=$ even. 

There are two sets of the complete basis for each parity. 
One is the natural basis with ${K}_{3}^{P}= {0}^{+}, {1}^{-}, {2}^{+}
, \cdots $, for odd $B$ and $K_{3}^{P} =
{\frac{1}{2}}^{+}, {\frac{3}{2}}^{-}, 
{\frac{5}{2}}^{+},\cdots $ for even $B$. 
Another is the unnatural basis with  ${K}_{3}^{P}= {0}^{-}, 
{1}^{+}, {2}^{-}, \cdots $, for odd $B$
and $K_{3}^{P} ={\frac{1}{2}}^{-}, {\frac{3}{2}}^{+}, 
{\frac{5}{2}}^{-}, \cdots $ for even $B$
The natural basis is given by 
\begin{eqnarray}
	&&{\phi}^{\it n}_\mu(\bm{x})= \nonumber \\
	&&\hspace{4mm}\left( 
	\begin{array}{c}
	 \displaystyle{\sum_{\alpha(0)}}f_{\alpha(0),\mu}
	 \bm{\Phi}_{\alpha(0)}(\bm{x},\uparrow_S)
	 +\displaystyle{\sum_{\alpha(1)}}f_{\alpha(1),\mu}
	 \bm{\Phi}_{\alpha(1)}(\bm{x},\downarrow_S) \\
	i \displaystyle{\sum_{\beta(0)}}g_{\beta(0),\mu}
	  \bm{\Phi}_{\beta(0)}(\bm{x},\uparrow_S)
	+i\displaystyle{\sum_{\beta(1)}}g_{\beta(1),\mu}
	  \bm{\Phi}_{\beta(1)}(\bm{x},\downarrow_S)
	\end{array}\right)\; {\chi}^I_{u}
	\nonumber \\
	&&+\left(
	\begin{array}{c}
	 \displaystyle{\sum_{\alpha(2)}}f_{\alpha(2),\mu}
	 \bm{\Phi}_{\alpha(2)}(\bm{x},\uparrow_S)
	+\displaystyle{\sum_{\alpha(3)}}f_{\alpha(3),\mu} 
	 \bm{\Phi}_{\alpha(3)}(\bm{x},\downarrow_S)
	\\
	 i\displaystyle{\sum_{\beta(2)}}g_{\beta(2),\mu}
	  \bm{\Phi}_{\beta(2)}(\bm{x},\uparrow_S)
	+i\displaystyle{\sum_{\beta(3)}}g_{\beta(3),\mu}
	  \bm{\Phi}_{\beta(3)}(\bm{x},\downarrow_S)
	\end{array}\right)\; {\chi}^I_{d}
	\label{expansion}
\end{eqnarray}
where
\begin{eqnarray*}
	&&\alpha (0) = \{n_r, n_z: {\rm odd}, 
	~\omega_0\equiv K_3-1/2-m_{\rm w}/2\}  \\
	&&\alpha (1) = \{n_r, n_z: {\rm even},
	\omega_1\equiv K_3+1/2-m_{\rm w}/2\}  \\
	&&\alpha (2) = \{n_r, n_z: {\rm even},
	\omega_2\equiv K_3-1/2+m_{\rm w}/2\}  \\
	&&\alpha (3) = \{n_r, n_z: {\rm odd},
	~\omega_3\equiv K_3+1/2+m_{\rm w}/2\}  \\
\end{eqnarray*}
and 
\begin{eqnarray*}
	&&\beta (0) = \{n_r, n_z: {\rm even}, 
	\omega_0\equiv K_3-1/2-m_{\rm w}/2\} \\
	&&\beta (1) = \{n_r, n_z: {\rm odd},
	~\omega_1\equiv K_3+1/2-m_{\rm w}/2\} \\
	&&\beta (2) = \{n_r, n_z: {\rm odd},
	~\omega_2\equiv K_3-1/2+m_{\rm w}/2\} \\
	&&\beta (3) = \{n_r, n_z: {\rm even}, 
	\omega_3\equiv K_3+1/2+m_{\rm w}/2\}. \\
\end{eqnarray*}
The unnatural basis $\phi_{\mu}^{(u)}$ is given by replacing, 
$\alpha \leftrightarrow \beta$ in Eq.~(\ref{expansion}).
By using the natural and unnatural basis, the eigenvalue problem
in Eq.~(\ref{eigenequation2}) can be reduced to a symmetric matrix  
diagonalization problem.  

Let us calculate the matrix elements of the hamiltonian below. 
For the kinetic term $\bm{\alpha} \cdot \bm{p}$ , we have 
\begin{eqnarray}
	&&\langle
	\bm{\Phi}_{\alpha(0)}|\bm{\sigma} \cdot \bm{p}|i\bm{\Phi}_{\beta'(0)}\rangle
	\nonumber 
	\\
	&&= \frac{1}{2 \pi} \int d^{3} x   {\phi}^{|\omega_0|}_{n_r}(\rho)
 	{\phi}_{n_z}(z) e^{-i\omega_0 \varphi } 
	\Bigl(\frac{\partial}{\partial z}\Bigr) 
	{\phi}^{|\omega_0'|}_{n_r'} (\rho) 
	{\phi}_{n_z'} (z) e^{i\omega_0'\varphi}
	\nonumber
	\\
	&&
	= {\delta}_{n_r n_r'}( {N_{n_z} N_{n_z'}}
	\sqrt{\alpha_z} n_z' \frac{1}{N^{2}_{n_{z}}} 
	{\delta}_{{n_z}{n_z'-1}}  
	-\frac{1}{2}{N_{n_z} N_{n_z'}}\sqrt{\alpha_z}
	\frac{1}{N^{2}_{n_{z}}} {\delta}_{{n_z} {n_z'+1}})~~\nonumber\\
	&&
	=\left\{ 
	\begin{array}{c} 
 	 \delta_{\omega_0 \omega_0'} {\delta}_{n_r n_r'} \frac{N_{n_z}'}{N_{n_z}} 
 	 \sqrt{\alpha_z} n_z' {\delta}_{{n_z}{n_z'-1}}\\
	 \delta_{\omega_0 \omega_0'} {\delta}_{{n_r} {n_r'}} (-\frac{1}{2}) \frac{N_{n_z'}}
	{N_{n_z}} \sqrt{\alpha_z}{\delta}_{n_z n_z'+1 }\,,
	\end{array}\right.\;
	\label{eq:83} \\
	&&\langle
	\bm{\Phi}_{\alpha(0)}|\bm{\sigma} \cdot \bm{p}|
	i\bm{\Phi}_{\beta'(1)}\rangle
	\nonumber
	\\
	&&= \frac{1}{2 \pi} \int d^{3} x
	{\phi}^{|\omega_0|}_{n_{r}}(\rho) {\phi}_{n_{z}}(z) 
	e^{-i\omega_0\varphi }
	e^{-i \varphi}\Bigl(\frac{\partial}{\partial \rho} 
	- \frac{i}{\rho} \frac{\partial}
	{\partial \varphi}\Bigr) {\phi}^{|\omega_1'|}_{n_r'} (\rho) \phi_{n_z'} (z) 
	e^{i\omega_1'\varphi }
	\nonumber
	\\
	&&=\left\{ 
	\begin{array}{c}
	 \delta_{n_z n_z'}\sqrt{\alpha_r} (\sqrt{n_r+\omega_0+1} 
	 \delta_{n_r n_r'} +\sqrt{n_r}\delta_{n_r-1 n_r'}) \\
	(\omega_0 \ge 0 :\omega_1'=\omega_0+1>0) \\
	-\delta_{n_z n_z'} \sqrt{\alpha_r}(\sqrt{n_r-\omega_0} 
	 \delta_{n_r n_r'} +\sqrt{n_r+1}
	 \delta_{n_r n_r'-1}). \\
	(\omega_0 < 0 :\omega_1'= \omega_0+1 \le 0)
	\end{array}\right.\;
	\label{eq:84} 
\end{eqnarray} 
In the natural basis, quantum numbers $(n_z,n_z')$ 
takes values $(1,2),(3,4), \cdots$ for the upper part 
and $(1,0),(3,2), \cdots$ for the lower part. 
In the unnatural basis, $(n_{z},{n'}_{z})=(0,1),(2,3), \cdots$ 
for the upper part and $(n_{z},{n'}_{z})=(2,1),(3,2), \cdots$ 
for the lower part. 

For the potential term  
$\beta M (\cos F(\rho,z)  + i {\gamma}_{5}
\bm{\tau} \cdot \hat{\bm{n}}_R \sin F(\rho,z))$\,,
we have 
\begin{eqnarray}
	&&\langle
	 \bm{\Phi}_{\alpha(0)}\chi^I_u| M \cos F(\rho,z)
	 |\bm{\Phi}_{\alpha'(0)}\chi^I_u \rangle 
	 \nonumber
	 \\
	&&= \int \rho d \rho dz M \cos F(\rho,z) 
	{\phi}^{|\omega_0|}_{n_{r}} (\rho) {\phi}_{n_{z}}(z)
	{\phi}^{|\omega_0|}_{{n'}_{r}} 
	(\rho) {\phi}_{{n'}_{z}} (z)\,, \\
	&&
	\langle \bm{\Phi}_{\alpha(0)}\chi^I_u| M i \bm{\tau}
	\cdot \hat{\bm{n}}_R \sin F(\rho,z)
	|i\bm{\Phi}_{\beta'(0)}\chi^I_u\rangle
	\nonumber
	\\
	&&= -\int \rho d \rho dz M \cos {\Theta}(\rho,z) \sin F(\rho,z) 
	{\phi}^{|\omega_0|}_{n_{r}} (\rho) {\phi}_{n_{z}}(z)
	  {\phi}^{|\omega_o'|}_{{n'}_{r}} 
	 (\rho) {\phi}_{{n'}_{z}} (z)\,,
	\label{eq:92} \\
	&&\langle \bm{\Phi}_{\alpha(0)}\chi^I_u|
	M i \bm{\tau} \cdot \hat{\bm{n}}_R\sin F(\rho,z)|
	\bm{\Phi}_{\beta'(2)}\chi^I_d \rangle \nonumber \\
	&&=-\int \rho d \rho dz M \sin {\Theta}(\rho,z)
	\sin F(\rho,z){\phi}^{|\omega_0|}_{n_{r}}(\rho)
	{\phi}_{n_{z}}(z){\phi}^{|\omega_2|}_{{n'}_{r}}(\rho)
	{\phi}_{{n'}_{z}} (z)\,.\label{eq:93}
\end{eqnarray}
Other elements can be calculated in the same manner.

To estimate the moments of inertia, {\it e.g.}, $\langle n|\lambda_a|m \rangle$ 
let us introduce the basis for the strange direction~\cite{weigel92}. 
The hamiltonian for the strange quark commutes with the total angular momentum
\begin{eqnarray}
J_3=L_3+\frac{1}{2}\sigma_3
\end{eqnarray}
because of the trivial construction of the $SU(3)$ chiral fields in Eq.~(\ref{embedding}).
As a result, the deformed basis has common form with the $SU(2)$ 
(see Eqs.~(\ref{deformed_basis})-(\ref{expansion})),
except in the quantum number in Eq.~(\ref{expansion}) as
\begin{eqnarray*}
	&&\alpha (0) = \{n_r, n_z: {\rm even}, 
	\omega_0\equiv J_3-1/2\}  \\
	&&\alpha (1) = \{n_r, n_z: {\rm odd},
	~\omega_1\equiv J_3+1/2\}  
\end{eqnarray*}
and 
\begin{eqnarray*}
	&&\beta (0) = \{n_r, n_z: {\rm odd}, 
	~\omega_0\equiv J_3-1/2\} \\
	&&\beta (1) = \{n_r, n_z: {\rm even},
	\omega_1\equiv J_3+1/2\}\,.
\end{eqnarray*}
The unnatural basis $\phi_{\mu}^{(u)}$ is given by replacing, 
$\alpha \leftrightarrow \beta$ in Eq.~(\ref{expansion}).

\section{Operation $\hat{R}$ to the Numerical Basis}
In this appendix, we present the construction of an operator $\hat{R}$
to the Kahana-Ripka basis in detail. Since it operates on the spherical harmonics, 
let us write the spherical harmonics in terms of the complex variables $z$ 
and their conjugate $\bar{z}$ as follows :
\begin{eqnarray}
z=\tan\frac{\theta}{2}e^{i\varphi}\,,
\end{eqnarray}
which can be related to the usual polar coordinates by  
\begin{eqnarray}
\cos \theta=\frac{1-|z|^2}{1+|z|^2}\,,~~\exp(i\varphi)=\pm \sqrt{\frac{z}{\bar{z}}}
\end{eqnarray}
According to the definition of the spherical harmonics \cite{sharmonics}, 
we obtain (up to $l\le 3$)
\begin{eqnarray}
&&Y_{11}=-\frac{1}{2}\sqrt{\frac{3}{2\pi}}\frac{2z}{1+|z|^2}\,,
~~Y_{10}=\frac{1}{2}\sqrt{\frac{3}{\pi}}\frac{1-|z|^2}{1+|z|^2}\,,
~~Y_{1-1}=\frac{1}{2}\sqrt{\frac{3}{2\pi}}\frac{2\bar{z}}{1+|z|^2}\,, \nonumber \\
\\
&&Y_{22}=\frac{1}{4}\sqrt{\frac{3\cdot 5}{2\pi}}\Bigl(\frac{2z}{1+|z|^2}\Bigr)^2\,,
~~Y_{21}=-\frac{1}{2}\sqrt{\frac{3\cdot 5}{2\pi}}\frac{2z(1-|z|^2)}{(1+|z|^2)^2}\,, \nonumber \\
&&Y_{20}=\frac{1}{4}\sqrt{\frac{5}{\pi}}\biggl(3\Bigl(\frac{1-|z|^2}{1+|z|^2}\Bigr)^2-1\biggr)\,,
\\
&&Y_{2-1}=\frac{1}{2}\sqrt{\frac{3\cdot 5}{2\pi}}\frac{2\bar{z}(1-|z|^2)}{(1+|z|^2)^2}\,,
~~Y_{2-2}=\frac{1}{4}\sqrt{\frac{3\cdot 5}{2\pi}}\Bigl(\frac{2\bar{z}}{1+|z|^2}\Bigr)^2\,, \nonumber \\
\nonumber \\
&&Y_{33}=-\frac{1}{8}\sqrt{\frac{5\cdot 7}{\pi}}\Bigl(\frac{2z}{1+|z|^2}\Bigr)^3\,,
~~Y_{32}=\frac{1}{4}\sqrt{\frac{3\cdot 5\cdot 7}{2\pi}}\frac{4z^2(1-|z|^2)}{(1+|z|^2)^3}\,,\nonumber \\
&&Y_{31}=-\frac{1}{8}\sqrt{\frac{3\cdot 7}{\pi}}\biggl(5\Bigl(\frac{1-|z|^2}{1+|z|^2}\Bigr)^2-1\biggr)
\frac{2z}{1+|z|^2}\,,\nonumber \\
&&Y_{30}=\frac{1}{4}\sqrt{\frac{7}{\pi}}\biggl(5\Bigl(\frac{1-|z|^2}{1+|z|^2}\Bigr)^2-3\biggr)
\frac{1-|z|^2}{1+|z|^2}\,, \\
&&Y_{3-1}=\frac{1}{8}\sqrt{\frac{3\cdot 7}{\pi}}\biggl(5\Bigl(\frac{1-|z|^2}{1+|z|^2}\Bigr)^2-1\biggr)
\frac{2\bar{z}}{1+|z|^2}\,,\nonumber \\
&&Y_{3-2}=\frac{1}{4}\sqrt{\frac{3\cdot 5\cdot 7}{2\pi}}\frac{4\bar{z}^2(1-|z|^2)}{(1+|z|^2)^3}\,,
~~Y_{3-3}=\frac{1}{8}\sqrt{\frac{5\cdot 7}{\pi}}\Bigl(\frac{2\bar{z}}{1+|z|^2}\Bigr)^3\,.
\nonumber
\end{eqnarray}
According to the (inverse) transformation of the $Z_2\times Z_2$ 
\begin{eqnarray}
z\to -\frac{1}{z}~~\equiv~~\theta\to \pi-\theta, \varphi\to \pi-\varphi\,,
\end{eqnarray}
the spherical harmonics are transformed via
\begin{eqnarray}
Y_{lm}(\theta,\varphi)\to Y_{lm}(\pi-\theta,\pi-\varphi)=(-1)^{l-m}Y_{l-m}(\theta,\varphi)\,.
\label{zrotation}
\end{eqnarray}
For the (inverse) tetrahedral transformation, we have 
\begin{eqnarray}
z\to i\, \frac{1-z}{1+z}\,,
\end{eqnarray} 
Thus the transformation of the spherical harmonics is given by  
\begin{eqnarray}
&&Y_{11}\to -\frac{i}{2}Y_{11}-\frac{i}{\sqrt{2}}Y_{10}-\frac{i}{2}Y_{1-1}\,, \nonumber \\
&&Y_{10}\to -\frac{i}{\sqrt{2}}Y_{11}+\frac{i}{\sqrt{2}}Y_{1-1}\,, \\
&&Y_{1-1}\to \frac{i}{2}Y_{11}+\frac{i}{\sqrt{2}}Y_{10}+\frac{i}{2}Y_{1-1}\,. 
\label{sphericaltrans1}
\nonumber \\
\nonumber \\
&&Y_{22}\to -\frac{1}{4}Y_{22}-\frac{1}{2}Y_{21}-\frac{1}{2}\sqrt{\frac{3}{2}}Y_{20}
 -\frac{1}{2}Y_{2-1}-\frac{1}{4}Y_{2-2}\,, \nonumber \\
&&Y_{21}\to \frac{i}{\sqrt{2}}Y_{22}+\frac{i}{\sqrt{2}}Y_{21}-\frac{i}{2}Y_{2-1}-\frac{i}{2}Y_{2-2}\,, \nonumber \\
&&Y_{20}\to \frac{1}{2}\sqrt{\frac{3}{2}}Y_{22}-\frac{1}{2}Y_{20}+\frac{1}{2}\sqrt{\frac{3}{2}}Y_{2-2}\,, \\
&&Y_{2-1}\to -\frac{i}{\sqrt{2}}Y_{22}+\frac{i}{\sqrt{2}}Y_{21}-\frac{i}{2}Y_{2-1}+\frac{i}{2}Y_{2-2}\,, \nonumber \\
&&Y_{2-2}\to -\frac{1}{4}Y_{22}+\frac{1}{2}Y_{21}-\frac{1}{2}\sqrt{\frac{3}{2}}Y_{20}\,,
 +\frac{1}{2}Y_{2-1}-\frac{1}{4}Y_{2-2}\,. 
\label{sphericaltrans2} 
\nonumber \\
\nonumber \\
&&Y_{33}\to \frac{i}{8}Y_{33}+\frac{\sqrt{6}i}{8}Y_{32}+\frac{\sqrt{15}i}{8}Y_{31}+\frac{\sqrt{5}i}{4}Y_{30}
+\frac{\sqrt{15}i}{8}Y_{3-1}+\frac{\sqrt{6}i}{8}Y_{3-2}+\frac{i}{8}Y_{3-3}\,, \nonumber \\
&&Y_{32}\to \frac{1}{4}\sqrt{\frac{3}{2}}Y_{33}+\frac{1}{2}Y_{32}+\frac{1}{4}\sqrt{\frac{5}{2}}Y_{31}
-\frac{1}{4}\sqrt{\frac{5}{2}}Y_{3-1}-\frac{1}{2}Y_{3-2}-\frac{1}{4}\sqrt{\frac{3}{2}}Y_{3-3}\,, \nonumber \\
&&Y_{31}\to -\frac{\sqrt{15}i}{8}Y_{33}-\frac{\sqrt{10}i}{8}Y_{32}+\frac{i}{8}Y_{31}+\frac{\sqrt{3}i}{4}Y_{30}
+\frac{i}{8}Y_{3-1}-\frac{\sqrt{10}i}{8}Y_{3-2}-\frac{\sqrt{15}i}{8}Y_{3-3}\,, \nonumber \\
&&Y_{30}\to -\frac{\sqrt{5}}{4}Y_{33}+\frac{\sqrt{3}}{4}Y_{31}-\frac{\sqrt{3}}{4}Y_{3-1}+\frac{\sqrt{5}}{4}Y_{3-3}\,,\\
&&Y_{3-1}\to \frac{\sqrt{15}i}{8}Y_{33}-\frac{\sqrt{10}i}{8}Y_{32}-\frac{i}{8}Y_{31}+\frac{\sqrt{3}i}{4}Y_{30}
-\frac{i}{8}Y_{3-1}-\frac{\sqrt{10}i}{8}Y_{3-2}+\frac{\sqrt{15}i}{8}Y_{3-3}\,, \nonumber \\
&&Y_{3-2}\to \frac{1}{4}\sqrt{\frac{3}{2}}Y_{33}-\frac{1}{2}Y_{32}+\frac{1}{4}\sqrt{\frac{5}{2}}Y_{31}
-\frac{1}{4}\sqrt{\frac{5}{2}}Y_{3-1}+\frac{1}{2}Y_{3-2}-\frac{1}{4}\sqrt{\frac{3}{2}}Y_{3-3}\,, \nonumber \\
&&Y_{3-3}\to -\frac{i}{8}Y_{33}+\frac{\sqrt{6}i}{8}Y_{32}-\frac{\sqrt{15}i}{8}Y_{31}+\frac{\sqrt{5}i}{4}Y_{30}
-\frac{\sqrt{15}i}{8}Y_{3-1}+\frac{\sqrt{6}i}{8}Y_{3-2}-\frac{i}{8}Y_{3-3}\,. 
\label{sphericaltrans3}
\nonumber 
\end{eqnarray}
The operation $\hat{R}$ to the basis becomes 
\begin{eqnarray}
\hat{R}|(lj)KM\rangle\equiv \hat{K} \sum_{j_3\tau_3}C^{KM}_{jj_3\frac{1}{2}\tau_3}
\Bigl(\sum_{m\sigma_3}C^{jj_3}_{lm\frac{1}{2}\sigma_3}
|lm \rangle' |\frac{1}{2}\sigma_3 \rangle \Bigr) |\frac{1}{2} \tau_3 \rangle\,,
\label{abasetrans}
\end{eqnarray}
where $|lm\rangle'$ representa the transformations of the spherical harmonics in Eqs.~(\ref{zrotation}) or 
(\ref{sphericaltrans1})-(\ref{sphericaltrans3}). Evaluationg $\hat{R}$ for each symmetric operation, 
one can get the final results in Eqs.~(\ref{basetransz})-(\ref{basetrans2}).

\newpage 

\begin{figure}

\caption{\label{fig:Sf} Spectra of the quark orbits 
of $B=2$ and $B=4$ with axially symmetric ansatz (\ref{ansatz}), 
as a function of the soliton size parameter $X$~\cite{komori04}.}
\end{figure}

\begin{figure}
\caption{\label{fig:Pf23} Contour plot of the profile functions
$F(\rho,z),\Theta(\rho,z)$, of $B=2,3$ with axial symmetry~\cite{komori04}.}
\end{figure}

\begin{figure}
\caption{\label{fig:Pf45} Contour plot of the profile functions
$F(\rho,z),\Theta(\rho,z)$, of $B=4,5$  with axial symmetry~\cite{komori04}.}
\end{figure}

\begin{figure}
\caption{\label{fig:Bd}Contour plot of the 
baryon number densities $b(\bm{x})$ [${\rm fm^{-3}}$] (\ref{bdensity_axial}) 
with axial symmetry~\cite{komori04}.}
\end{figure}

\begin{figure}
\begin{center}
\caption{\label{fig:profile} Self-consistent profile 
functions for $B=3-9,17$ in the rational map ansatz calculations~\cite{sawado04}.}
\end{center}
\end{figure}
\begin{center}
\begin{figure}
\caption{\label{fig:bdensity} Surface plot of the baryon 
number densities $b(\bm{x})$ (\ref{baryon_density}) for $B=3-9$ and 
the excited states $B=5^*,9^*$~\cite{sawado04}.}
\end{figure}
\end{center}

\begin{figure}
\caption{\label{fig:spectrum} Valence quark spectra for $B=1-9,17$~\cite{sawado04}.}
\end{figure}
\begin{figure}
\caption{\label{fig:bdrs} Angular averaged baryon densities 
of $i$th valence quarks $\rho^{(i)}(r)$ of $B=3-9,17$, 
with the occupation number and the eigenvalue (in MeV)~\cite{sawado04}.}
\end{figure}

\begin{figure}
\caption{\label{fig:bdensityb2} Surface plot of the baryon number 
densities for $B=2$ configurations (\ref{rmab2})~\cite{tanaka04}.}
\end{figure}
\begin{figure}
\caption{\label{fig:energyb2} Total energies of the $B=2$ configurations \cite{tanaka04}.}
\end{figure}

\begin{figure}[htbp]
\caption{\label{fig:su3_spectra}The dibaryon spectra~\cite{sawado02}.}
\end{figure}


\begin{thebibliography}{qq}
\bibitem{freedman93} S.\ J.\ Freedman {\it et al}., Phys.\ Rev.\ C {\bf 48}, 
            1864 (1993); 
\bibitem{gross92} F.\ Gross and S.\ Luiti, Phys.\ Rev.\ C {\bf 45} 1374 (1992).
\bibitem{thomas94} W.\ Melnitchouk, A.\ W.\ Schreiber, and A.\ W.\ Thomas, 
            Phys.\ Rev.\ D {\bf 49}, 1183 (1994).
\bibitem{umnikov96} A.\ Yu.\ Umnikov, F.\ C.\ Khanna and L.\ P.\ Kaptari, 
            Phys.\ Rev.\ C {\bf 53}, 377 (1996).
\bibitem{carlson95} C.\ E.\ Carlson and K.\ E.\ Lassila, Phys.\ Rev.\ C 
            {\bf 51}, 364 (1995).
\bibitem{diakonov97} D. Diakonov, V. Petrov and M. Polykov, Z. Phys. A 
{\bf 359}, 305 (1997).
\bibitem{nakano03} T. Nakano {\it et al.}, Phys. Rev. Lett. {\bf 91}, 012002
 (2003).
 \bibitem{diakonov88}
D. I. Diakonov, V. Yu. Petrov, and P. V. Pobylitsa, 
Nucl. Phys. B {\bf 306}, 809 (1988).
\bibitem{reinhardt88}
H. Reinhardt and R. W\"unsch , Phys. Lett. B {\bf 215}, 577 (1988).
\bibitem{meissner89}
Th. Meissner, F. Gr\"ummer, and K. Goeke, Phys. Lett. B
{\bf 227}, 296 (1989).
\bibitem{report96} For detailed reviews of the model see: \\
R.\ Alkofer, H.\ Reinhardt and H.\ Weigel, Phys.\ Rept.\ {\bf 265}, 139 (1996);\\
 Chr.\ V.\ Christov, A.\ Blotz, H.-C.Kim, P.\ Pobylitsa, T.\ Watabe, Th.\ Meissner, 
E.\ Ruiz Arriola, K.\ Goeke, Prog.\ Part.\ Nucl.\ Phys.\ {\bf 37}, 91 (1996).
\bibitem{wakamatsu91}
M. Wakamatsu and H. Yoshiki, Nucl. Phys. A {\bf 524}, 561 (1991).
\bibitem{sawado98}
N. Sawado and S. Oryu, Phys. Rev. C {\bf 58}, R3046 (1998).
\bibitem{sawado00}
N. Sawado, Phys. Rev. C {\bf 61}, 65206 (2000).
\bibitem{sawado02}
N. Sawado, Phys. Lett. B {\bf 524}, 289 (2002).
\bibitem{braaten90}
E. Braaten, S. Townsend and L. Carson, 
Phys. Lett. B {\bf 235}, 147 (1990).
\bibitem{sutcliffe97}
R.A.Battye and P.M.Sutcliffe, 
Phys. Rev. Lett. {\bf 79},363 (1997).
\bibitem{manton98}
C. J. Houghton, N. S. Manton and P. M. Sutcliffe, 
Nucl. Phys. B {\bf 510}, 507 (1998).
\bibitem{sawado02t}
N. Sawado and N. Shiiki,
Phys. Rev. D {\bf 66}, 011501 (2002).
\bibitem{sawado04}
Nobuyuki Sawado and Noriko Shiiki, 
{\it hep-ph/0402084}.
\bibitem{komori04} S. Komori, N. Sawado and N. Shiiki, 
Ann. Phys. {\bf 311}, 1 (2004).
\bibitem{irwin}
P. Irwin, Phys. Rev. D {\bf 61}, 114024 (2000).
\bibitem{grigoriev}
D. Y. Grigoriev, P. M. Sutcliffe and D. H. Tchrakian,
Phys. Lett. B {\bf 540}, 146 (2002).
\bibitem{jaffe77} R.\ L.\ Jaffe, Phys.\ Rev.\ Lett.\ {\bf 38}, 195 (1977); 
{\bf 38}, 1617(E) (1977).
\bibitem{dhar85}
A.Dhar, R.Shankar, S.R.Wadia, 
Phys. Rev. D {\bf 31}, 3256 (1985).
\bibitem{ebert86}
D. Ebert, H. Reinhardt, 
Nucl.Phys. B {\bf 271}, 188,(1986).
\bibitem{aitchison85}
I. Aitchison, C. Fraser, E. Tudor and J. Zuk,  
Phys. Lett. B {\bf 165}, 162 (1985).
\bibitem{kahana84}
S. Kahana, G. Ripka, and V. Soni, Nucl. Phys. A {\bf 415}, 
351 (1984); S. Kahana and G. Ripka, \textit{ibid}, A {\bf 429}, 462 (1984).
\bibitem{balachandran98}
A. P. Balachandran and S. Vaidya,
Int. J. Mod. Phys. A {\bf 14}, 445 (1999),
{\it hep-th/9803125}.
\bibitem{rajaraman}
R.Rajaraman, {\it Solitons and Instantons}
(North-Holland Physics Publishing, Amsterdam, 1987).
\bibitem{reinhardt89}
H. Reinhardt, Nucl. Phys. A {\bf 503}, 825 (1989).
\bibitem{schwinger51}
J. Schwinger, Phys. Rev. {\bf 82}, 664 (1951).
\bibitem{manton87}
N. S. Manton, Phys. Lett. B {\bf 192,} 177 (1987). 
\bibitem{kopeliovich87}
V. B. Kopeliovich and B. E. Stern, Pis'ma Zh. \'Eksp. 
Teor. Fiz. {\bf 45,} 165 (1987) [JETP Lett. {\bf 45}, 203 (1987)].
\bibitem{verbaashot87}
J. Verbaarschot, Phys. Lett. B {\bf 195}, 235 (1987).
\bibitem{weigel86}
H. Weigel, B. Schwesinger and G. Holzwarth, Phys. Lett. B 
{\bf 168}, 321 (1986).
\bibitem{braaten88}
E. Braaten and L. Carson, Phys. Rev. D {\bf 38}, 3525 (1988).
\bibitem{battye01}
R. A. Battye and P. M. Sutcliffe, 
Rev. Math. Phys. {\bf 14}, 29, (2002).
\bibitem{bohr}
A. Bohr and B. Mottelson, {\it Nuclear structure, Vol.II}
(World Scientific Publishing Co. Pte. Ltd, Singapore, 1998).
\bibitem{manton00}
N. S. Manton and B. M. A. G. Piette, 
Prog. Math. {\bf 201}, 469 (2001);{\it hep-th/0008110}.
\bibitem{moussallam91}
B.\ Moussallam and D.\ Kalafatis, Phys. Lett. B {\bf 272}, 196 (1991).
\bibitem{moussallam93}
B.\ Moussallam, Ann. Phys. {\bf 225}, 264 (1993).
\bibitem{holzwarth}
G.\ Holzwarth and H.\ Walliser, Nucl. Phys. A {\bf 587}, 721 (1995).
\bibitem{weigel95}
H.\ Weigel, R.\ Alkofer and H.\ Reinhardt, Nucl. Phys. A {\bf 582}, 484 (1995).
\bibitem{scholtz}
F.\ G.\ Scholtz, B.\ Schwesinger and H.\ B.\ Geyer, 
Nucl. Phys. A {\bf 561}, 542 (1993).
\bibitem{manton82}
N. S. Manton, Phys. Lett. B {\bf 110}, 54 (1982). 
\bibitem{manton77}
N. S. Manton, Nucl. Phys. B {\bf 126}, 525 (1977). 
\bibitem{bogomol'nyi}
E. B. Bogomol'nyi, Sov. J. Phys. {\bf 24}, 97 (1977). 
\bibitem{tanaka04}
N. Sawado, N. Shiiki and Y. Tanaka, {\it unpublished}.
\bibitem{gibbons86}
G. W. Gibbons and N. S. Manton, Nucl. Phys. B {\bf 274}, 183 (1986).
\bibitem{leese90}
Robert Leese, Nucl. Phys. B {\bf 344}, 33 (1990).
\bibitem{dudek}
J. Dudek, A. Go\'zd\'z, N. Schunck, and M. Mi\'skiewicz, 
Phys. Rev. Lett. {\bf 88}, 252502 (2002).
\bibitem{hamermesh}
M. Hamermesh, {\it Group theory and its application to physical
problems} (Dover Publications, Inc., New York, 1989).
\bibitem{koepf89}
W. Koepf and P. Ring, Nucl. Phys. A {\bf 493}, 61 (1989).
\bibitem{biedenharn85}
L. C. Biedenharn, Y. Dothan and 
M. Tarlini Phys. Rev. D {\bf 31,} 649 (1985).
\bibitem{goeke91} K. Goeke, A. Z. G\'orski, F. Gr\"ummer, Th. Meissner, 
H. Reinhardt, and R. W\"unsch, Phys. Lett. B {\bf 256}, 321 (1991).
\bibitem{fr} D. Finkelstein and J. Rubinstein, J. Math. Phys. {\bf 9}, 1762 (1968).
\bibitem{krusch} S. Krusch, Ann. Phys. {\bf 304}, 103 (2003);{\it hep-th/0210310}.
\bibitem{blotz93} A.\ Blotz, D.\ Diakonov, K.\ Goeke, N.\ W.\ Park, V.\ Petrov
and P.\ V.\ Pobylitsa, Nucl.\ Phys.\ A {\bf 555}, 765 (1993).
\bibitem{wakamatsu96} M.\ Wakamatsu, Phys,\ Rev,\ D {\bf 54}, 2161 (1996).  
\bibitem{weigel92} H.\ Weigel, R.\ Alkofer and H.\ Reinhardt, Nucl.\ Phys.\ B {\bf 387}, 638 (1992).
\bibitem{kopeliovich90} V.\ B.\ Kopeliovich, B.\ Schwesinger and B.\ E.\ Stern, Nucl.\ Phys.\ A {\bf 549}, 485 (1992).
\bibitem{toyota87} N.\ Toyota, Prog.\ Theor.\ Phys.\ {\bf 77}, 688 (1987).
\bibitem{kaeding96} T.\ A.\ Kaeding, H.\ T.\ Williams, Comput.\ Phys.\ Commun. {\bf 98}, 398 (1996).
\bibitem{gambhir} 
Y. K. Gambhir, P. Ring, A. Thimet, Ann. Phys. {\bf 198}, 132 (1990).
\bibitem{kunz88} J.\ Kunz and P.\ J.\ Mulders, Phys.\ Lett.\ B {\bf 215}, 449 (1988).
\bibitem{blotz96} A.\ Blotz, M.\ Praszalowicz and K.\ Goeke, Phys.\ Rev.\ D {\bf 53}, 485 (1996).
\bibitem{sharmonics}
Milton Abramowitz and Irene A. Stegun,
{\it Handbook of Mathematical Functions} 
(Dover Publications, New York, 1968). 
\end{thebibliography}
\end{document}